\documentclass[prc,twocolumn,twoside,amsmath,amsfonts,amssymb]{revtex4}

\usepackage{graphicx}

\newcommand \Section [1] { \section {\boldmath #1} }
\newcommand \Subsection [1] { \subsection {\boldmath #1} }

\newcommand \etal {\emph{et al.}}
\newcommand \nb [1] {\mbox {$ #1 $}}

\newcommand \bk { \nonumber \\ }

\newcommand \ds \displaystyle
\newcommand \ts \textstyle

\newcommand \mn \mathnormal
\newcommand \mr \mathrm
\newcommand \ma \mathsf
\newcommand \cl \mathcal
\newcommand \vc \boldsymbol

\newcommand \ba [2] { #1 ^{\phantom\dagger} _{ #2 } }
\newcommand \sq [2] { #1 ^{ \mr 2 } _{ #2 } }
\newcommand \sr [2] { #1 ^{ \mr { 1/2 } } _{ #2 } }
\newcommand \ir [2] { #1 ^{ \mr { -1/2 } } _{ #2 } }
\newcommand \iv [2] { #1 ^{ \mr {-1} } _{ #2 } }
\newcommand \hc [2] { #1 ^{ \mn \dagger } _{ #2 } }
\newcommand \cc [2] { #1 ^{ \mn \ast } _{ #2 } }
\newcommand \tp [2] { #1 ^{ \mn T } _{ #2 } }
\newcommand \rv \overline
\newcommand \nr [1]
  { \makebox [0 pt] { \scriptsize \phantom {$\rv #1$} } #1 }   
\newcommand \cj \tilde
\newcommand \rs { \makebox [0 pt] { \phantom {$\big|$} } }
\newcommand \iotakappa
  { \makebox [0 pt] { \scriptsize \phantom {$\lambda$} } \iota \kappa } 
\newcommand \lambdamus { \text {\scriptsize$ \begin{matrix}
    ( \mu  , \lambda ) \in \cl S \  \text{or} \\
    ( \lambda ^{-1} , \mu ^{-1} ) \in \cl S
  \end{matrix} $} }
\newcommand \phz {\phantom{0}}
\newcommand \phzz {\phantom{00}}
\newcommand \phd {\phantom{.000}}

\newcommand \ab [1] { { | #1 | } }
\newcommand \ke [1] { { | #1 \rangle } }
\newcommand \me [3] { { \langle #1 | #2 | #3 \rangle } }
\newcommand \av [1] { { \langle #1 \rangle}}
\newcommand \Av [1] { { \bigl\langle #1 \bigr\rangle } }
\newcommand \AV [1] { { \Bigl\langle #1 \Bigr\rangle } }
\newcommand \rb [2] { { \bigl( #1 \bigr) _{ #2 } } }

\newcommand \tbt [4] { \text {\scriptsize$ \begin{pmatrix}
    \ma { #1 } & \ma { #2 } \\ \ma { #3 } & \ma { #4 } 
  \end{pmatrix} $} }
\newcommand \tbo [2] { \text {\scriptsize$ \begin{pmatrix}
    \ma { #1 } \\ \ma { #2 }
  \end{pmatrix} $} }

\newcommand \hf { \tfrac 1 2 }
\newcommand \mhf { \mr { 1/2 } }
\newcommand \mmhf { \mr { -1/2 } }
\newcommand \rt { \sqrt 2 }
\newcommand \rr { \tfrac 1 { \sqrt 2 } }
\newcommand \jm \jmath

\newcommand \sy { \text s }
\newcommand \ay {\text a}
\newcommand \n {\text n}
\newcommand \p { \text p }
\newcommand \nn {\text {nn} }
\newcommand \np {\text {np} }
\newcommand \nnpp {\text {nn+pp} }
\newcommand \va { \text v }
\newcommand \tk { \text k }
\newcommand \pair { \text {pair} }
\newcommand \ph { \text {ph} }
\newcommand \HB { \text {HB} }
\newcommand \RPA { \text {RPA} }

\newcommand \col { \text {col} }
\newcommand \tr { \text {tr} }
\newcommand \twoG { \text {$ 2 G $} }
\newcommand \dMeV { \text {MeV} }
\newcommand \MeV { \; \dMeV }

\newcommand \sfx { { \ma x } }
\newcommand \sfy { { \ma y } }
\newcommand \sfz { { \ma z } }
\newcommand \sfn { { \ma n } }
\newcommand \sfp { { \ma p } }

\newcommand \pdv[2] { \frac { \partial #1 } { \partial #2 } }

\begin{document}

\title {Pairing theory of the symmetry energy}

\author {K. Neerg\aa rd}

\affiliation {Fjordtoften 17, 4700 N\ae stved, Denmark}

\email {kai@kaineergard.dk}

\begin{abstract}

A model is investigated which displays a picture of the symmetry energy
as an energy of rotation in isospace of a Cooper pair condensate,
briefly ``superfluid isorotation''. The Hamiltonian is isobarically
invariant and has a one- and a two-nucleon term, where the two-nucleon
interaction is composed of an isovector pairing force and an interaction
of isospins. It is analyzed in the Hartree-Bogolyubov plus Random Phase
approximation. The Hartree-Bogolyubov energy minus Lagrangian multiplier
terms proportional to the number of valence nucleons and the $z$
component of the isospin is shown to be locally minimized by a product
of neutron and proton Bardeen-Cooper-Schrieffer states. The equations of
the Random Phase Approximation (RPA) can be reduced to independent
equations for two-neutron, two-proton, and neutron-proton quasiparticle
pairs. In each of these spaces, they have a Nambu-Goldstone solution due
to the global gauge invariance and isobaric invariance of the
Hamiltonian. Except for the Nambu-Goldstone solutions, the RPA solutions
are independent of the strength of the isospin interaction. If, in one
space, the pertinent single-nucleon spectrum has a particle-hole
symmetry, the RPA solutions are twofold degenerate except for the
Nambu-Goldstone solution and one more solution. In an idealized case of
infinitely many equidistant single-nucleon levels, the one-nucleon term
in the Hamiltonian and the isospin interaction contribute terms in the
symmetry energy quadratic in the isospin $T$. The pairing force and the
two-neutron and two-proton RPA correlation energies do not contribute.
The contribution of the neutron-proton correlation energy is dominated
by the Nambu-Goldstone solution, which gives a linear term that makes
the total symmetry energy proportional to $T(T+1)$. The rest of this
contribution is negative and can be written as the difference of two
terms of the form $\sqrt { ( a T ) ^2 + b ^2 } - b$. Observations
reported from Skyrme force calculations are discussed in the light of
these results. Calculations with deformed Woods-Saxon single-nucleon
levels give results similar to those of the idealized case. In
calculations for the mass numbers $A = 56$ and $A = 100$ with spherical
Woods-Saxon levels, the promotion of nucleons across magic gaps in the
single-nucleon spectrum and the onset of superfluidity with the
departure from magicity give rise to large linear terms in the symmetry
energy. The calculations with Woods-Saxon single-nucleon levels
reproduce surprisingly well the empirical symmetry energy. An
experimental signature of superfluid isorotation is discussed.

\end{abstract}


\maketitle

\Section {\label{intr}Introduction}

The concept of a symmetry energy originates in Weizs\"acker's early
attempt~\cite{We} to construct a formula for the nuclear binding energy.
Guided by Majoranas ideas~\cite{Ma} as to the nature of the internucleon
force, Weizs\"acker suggests an expression for the binding energy of a
doubly even nucleus which, except for an electrostatic term, is
symmetric in the numbers $N$ and $Z$ of neutrons and protons. If
$B_\sy ( N , Z )$ is the symmetric part of such an expression, the
symmetry energy is $B_\sy ( A/2 , A/2 ) - B_\sy( N , Z )$ with
\nb{A = N + Z}. As discussed, for example, by Bohr and
Mottelson~\cite{BoMo}, it carries information on basic aspects of the
nuclear structure. Models of the nucleosynthesis in core-collapse
supernovae rely on estimates of the masses of nuclei inaccessible to
experiment. The accuracy of such estimates depends on a valid
understanding of the origin of the symmetry energy, and data on the
abundancies of nuclides in the nearby universe may in turn constrain
nuclear models, as discussed in recent reviews by Arnould and
Goriely~\cite{ArGo}, and Arnould, Goriely, and Takahashi~\cite{ArGoTa}.
In infinite nuclear matter, the symmetry energy per nucleon is a
function of the nucleon density. The form of this function has a bearing
on the surface structure of finite nuclei, the structure of neutron
stars, and the dynamics of heavy-ion reactions. This research was
reviewed recently by Li, Chen, and Ko~\cite{LiChKo}.

Because $N-Z$ is twice the eigenvalue $M_T$ of the $z$ component $T_z$
of the isospin $\vc T$ and most nuclear ground states are approximate
eigenstates of $\vc T^2$ with the eigenvalue $T(T+1)$ given by
$T = \ab{M_T}$, the symmetry energy may be conceived as the
$T$-dependent part of the nuclear ground state energy in the limit of
isobaric invariance. Much theoretical and analytical work, reviewed in
Sect.~\ref{rev}, aims at describing its dependence on $T$. The empirical
evidence seems, at least, compatible with the conjecture that, in a
first approximation, the symmetry energy is proportional to the Casimir
invariant of the isospin algebra, $T(T+1)$.

In two previous brief articles~\cite{Ne}, I discuss a schematic,
micoscopic model which gives this $T$ dependence approximately for low
$T$. It is inspired by the Goswami's~\cite{Gw} observation that the
potential of interaction, in a superfluid nucleus, of the individual
nucleons with the condensate of Cooper~\cite{Co} pairs, in the following
referred to briefly as the ``pair potential'', is not only nondiagonal
in $N$ and $Z$ but also isobarically noninvariant. Frauendorf and
Sheikh~\cite{FrSh} point out that the symmetry energy may be conceived
accordingly as an energy of rotation of the condensate in isospace. An
expression for the symmetry energy proportional to $T(T+1)$ is then
analogous to the well known expression for the energy levels of a
quantal rotor. It should be stressed that it is, in this picture, the
entire symmetry energy that is proportional to $T(T+1)$ and not only the
contribution from the two-nucleon term in the Hamiltonian. In this
respect the isorotational picture differs basically from several models
reviewed in Sect.~\ref{rev}. For brevity, I call a rotation in isospace
of a Cooper pair condensate ``superfluid isorotation''.

The model introduced in Ref.~\cite{Ne} is designed to display this
physics. It thus involves valence nucleons obeying a schematic
Hamiltonian with a one- and a two-nucleon term which conserve the number
$A_\va$ of valence nucleons and the isospin. To calculate states with
arbitrary $A_\va$ and $T$ by a minimization, Lagrangian multiplier terms
proportional to $\hat A_\va$ and $T_z$ are subtracted from the
Hamiltonian, where $\hat A_\va$ is the operator with the eigenvalues
$A_\va$. This is equivalent to imposing neutron and proton chemical
potentials. The Hamiltonian minus Lagrangian multiplier terms is treated
in the Hartree-Bogolyubov plus Random Phase approximation, which is
known to lead to a separation of the collective degrees of freedom
associated with spontaneously broken symmetries of a many-body system.
Although the Hamiltonian is schematic, the principles of its treatment
are thus general and may be applied to any energy functional of a
Bogolyubov quasinucleon vacuum provided this functional is invariant
under global gauge transformations and isobaric transformations.

The two-nucleon interaction of the model has a pairing and a
particle-hole part. When the latter is omitted, the theory is for a
spherical nucleus equivalent to that of Ginoccio and
Wesener~\cite{GiWe}. Some of the results derived below are known from
their study. However, the present methods are quite different from
theirs. The relation of the present work to that of Ginoccio and Wesener
is discussed on the way in Sect.~\ref{the}.

In the present article, my model is analyzed to a much greater depth
than in Ref.~\cite{Ne}. Its mathematical structure is discussed in
detail, its symmetries are explored, and the formulas used in the
calculations are given explicitly. Furthermore, the calculations have
been considerably extended. In the idealized case of equidistant
single-nucleon levels considered in Ref.~\cite{Ne}, the number of such
levels has been enlarged by a factor more than 40 in order to rule out
any spurious effect of the its finiteness. These calculations have been
carried out for parameters appropriate for different mass numbers, and
the results are described in a more general form than previously. A
deviation from a linear $T$ dependence of the contribution to the
symmetry energy from the correlation energy calculated in the Random
Phase Approximation (RPA) has been traced to the correlations of
neutron-proton quasiparticle pairs and its origin understood. Finally,
in order to approach a description of actual nuclei, calculations with
Woods-Saxon single-nucleon levels have been carried out for several
isobaric chains.

To set up a background for the present work, I review in Sect.~\ref{rev}
previous theories of the symmetry energy dating back to the era of the
birth of nuclear physics in the 1930ies. Certain aspects of some early
models are elucidated in an appendix. The present empirical evidence as
to the $T$ dependence of the symmetry energy is also discussed in
Sect.~\ref{rev}. The formalism is then developed in Sect.~\ref{the},
and the calculations described and discussed in Sect.~\ref{cal}. After
some brief remarks in Sect.~\ref{sig} on the issue of an experimental
signature of superfluid isorotation, the study is summarized in
Sect.~\ref{sum}.

\Section {\label{rev}Theory and phenomenology of the symmetry energy.
          A review}

Bethe and Bacher~\cite{BeBa} introduce the assumption, which has since
then been common in the literature, that the symmetry energy depends
quadratically on $N-Z$. This means, in terms of isospin, that it is
proportional to $T^2$. A different $T$ dependence is derived
theoretically by Wigner~\cite{Wi}. He assumes that nucleons interact by
a two-body force and makes the, now obsolete, assumption that this force
is invariant under arbitrary unitary transformations of the nucleonic
spin and isospin. This implies, in particular, that any exchange force
must be of the Majorana type. Wigner then infers that the two-nucleon
force gives a contribution to the symmetry energy equal to $T(T+4)$
times a factor which he supposes depends only weakly on $T$. If this
factor is constant, the contribution of the two-nucleon force to the
symmetry energy has a term linear in $T$ besides the quadratic one. This
is the celebrated ``Wigner term''. Wigner's derivation is easily redone
with the $SU(4)$ group of unitary transformations of the nucleonic spin
and isospin replaced with the $SU(2)$ group of isobaric transformations.
As shown in the appendix, a factor $T(T+1)$ then replaces $T(T+4)$. A
term in the symmetry energy linear in $T$ gives rise to a cusp at
\nb{N = Z} in the curve of masses along an isobaric chain. Such cusps
are found, in fact, in an analysis of measured masses by Myers and
Swiatecki~\cite{MySwWig}.

Wigner estimates the contribution of the nucleon kinetic energy by the
Fermi gas model, which gives a leading term proportional to $T^2$. In
his model, the contributions of the one- and two-nucleon terms in the
Hamiltonian thus depend differently on $T$. This is true for most models
with a Hamiltonian composed of a one- and a two-nucleon term.

In the framework of the spherical shell model, and assuming conservation
of isospin and seniority, Talmi and Unna~\cite{TaUn} show that, for
nuclei whose valence nucleons occupy a single $j$-shell, the symmetry
energy is proportional to $T(T+1)$. In the spherical shell model, the
single-nucleon term in the Hamiltonian has the role played in the Fermi
gas model by the nucleon kinetic energy. When all valence nucleons
occupy the same $j$-shell, the sum of their single-nucleon energies is
constant for a given mass number, so the entire symmetry energy stems
from the residual two-nucleon interaction. While the result of Talmi and
Unna thus agrees superficially with that of Wigner's argument applied to
the isobaric $SU(2)$, it is shown in the appendix to imply that the
basic assumption of the latter, namely that the average interaction
energy in two-nucleon states with a definite symmetry in position and
spin is independent of $T$, does not hold for the isobarically invariant
isovector pairing force acting in a single $j$-shell.

In isobarically invariant shell model calculations for \nb{A = 24} and
\nb{A = 48}, Satu\l a~\etal~\cite{Sa*} find that omitting the
interaction of isoscalar nucleon pairs essentially eliminates the
deviation from a quadratic $T$ dependence of the symmetry energy. Since
the \nb{A = 48} nuclei whose calculated binding energies are analysed in
Ref.~\cite{Sa*} belong to the $1f_{7/2}$ shell, this seems to indicate,
in view of the result of Talmi and Unna, that the interaction of
isovector pairs used in the calculation has a major component which does
not conserve seniority. In line with an earlier study by
Brenner~\etal~\cite{Br*}, Satu\l a~\etal\ also consider a certain linear
combination of the binding energies of nuclei differing in $N$ and $Z$
by small numbers. These linear combination are constructed so as to
filter out a posible enhancement of the binding energy for \nb{N = Z}.
Both Brenner~\etal\ and Satu\l a~\etal\ find that, for a range of
\emph{sd} and \emph{fp} shell nuclei, their linear combinations vanish
essentially when the interaction of isoscalar pairs is turned off in
their shell model calculations. They hence infer that the observed
deviation from a quadratic $T$ dependence of the symmetry energy stems
from this part of the two-nucleon interaction.

Myers and Swiatecki~\cite{MySwWig} fit the measured nuclear masses with
a formula where the deviation from a quadratic dependence of the
symmetry energy on $N - Z$ decreases exponentially with $\ab{N - Z}$.
More recently, Myers~\cite{My} suggests that a term in the symmetry
energy linear in $\ab{N - Z}$ could arise because two nucleons in
identical states of orbital motion are more strongly bound than other
nucleon pairs. He assumes that the single-nucleon levels are fourfold
degenerate and the states of each level differ only by the directions of
their spins and isospins. Thus he neglects the single-nucleon spin-orbit
potential. His independent-nucleon model has, in fact, Wigner's $SU(4)$
symmetry. When $N$ neutrons and $Z$ protons occupy the lowest levels in
a potential well, and $N$ and $Z$ are even, the number of pairs in
identical states of orbital motion is then equal to $3A/2 - \ab{N - Z}$.
In the presense of a single-nucleon spin-orbit potential, this is the
number of pairs of nucleons in identical or time reversed states of
orbital motion and spin. Jensen, Hansen, and Jonson~\cite{JeHaJo} work
out a version of Myers's argument which maintains the $SU(4)$ symmetry
of the independent-nucleon model but includes exchange terms in the
assumed delta-force interaction of the nucleons. In Ref.~\cite{MySwTom},
Myers and Swiatecki explain the extra binding of nucleon pairs in
identical states of orbital motion by the congruence of the nodal
surfaces of the single-nucleon wave functions. They call the resulting
term in the nuclear binding energy, accordingly, a ``congruence''
energy. In Ref.~\cite{MySwCon}, these authors consider only pairs of a
neutron and a proton and give for the number of particularly strongly
bound pairs of this kind the expression $(A - \ab{N - Z})/2$, which is
the number of neutron-proton pairs in identical states of orbital motion
and spin. Since a congruence of nodal surfaces requires that the
stationary states of a nucleon have unique orbital wave functions, and
thus that the single-nucleon spin-orbit potential is neglected, counting
only nucleon pairs with parallel spins seems, however, inconsistent with
the congruence picture.

A contribution to the symmetry energy proportional to $T(T+1)$ is
pointed out by Bohr and Mottelson~\cite{BoMo} to arise from the
separable particle-hole interaction which, in the Hartree approximation,
generates a term in the single-nucleon potential proportional to $\vc T
\cdot \vc t$, where $\vc t$ is the single-nucleon isospin. Satu\l a and
Wyss~\cite{SaWySym} study the symmetry energy in Hartree-Fock
calculations with Skyrme forces where they ensure isobaric invariance by
omitting the Coulomb force and assuming equal neutron and proton masses.
They find that when the isospin-dependent parts of the Skyrme forces are
omitted, the symmetry energy is roughly proportional to $T^2$ with $T$
taken here equal to $M_T$. The isospin-dependent parts give additional
contributions which are nearly proportional to \nb{T(T+1)}. They thus
act similarly to the separable particle-hole interaction of Bohr and
Mottelson. The quadratic $T$ dependence of the remainder of the symmetry
energy may be understood to result from a redistribution of the nucleons
on their self-consistent energy levels. Fig.~4 of Ref.~\cite{SaWySym}
shows that when a Bardeen-Cooper-Schrieffer~\cite{BaCoSc} pairing term
is added to the energy functional, the contribution of the
isospin-dependent forces to the symmetry energy is no longer
proportional to $T(T+1)$, where $T$ is now taken equal to $\av {T_z}$. A
roughly quadratic dependence on $T$ can be inferred from the article's
Figs.~2 and~4. This is explained in Sect.~\ref{skm} below.

In a model with nucleons in a deformed potential and a pairing force
acting on both isoscalar and isovector nucleon pairs, Satu\l a and
Wyss~\cite{SaWyWig} find that an approximate particle number projection
allows isoscalar and isovector components of the pair potential to
coexist. In calculations for doubly even nuclei in the \emph{fp} shell,
the isoscalar component turns out to be for certain values of the
coupling constants particularly large for \nb{N \approx Z}, which leads
to an enhanced binding of such nuclei. The authors therefore suggest
that the Wigner term arises from isoscalar pairing. In
Ref.~\cite{SaWyCra}, they consider an isoscalar pair potential with a
different structure. Without particle number projection, Civitarese,
Reboiro, and Vogel~\cite{CiReVo} obtain a similar enhanced binding for
$N \approx Z$ with an isovector pairing force which acts with a larger
strength on neutron-proton than on neutron and proton pairs and is thus
isobarically noninvariant.

The empirical evidence as to the $T$ dependence of the symmetry energy
is ambiguous. Thus in the recent analyses by Royer and
Gautier~\cite{RoGa}, Royer~\cite{Ro}, Kirson~\cite{KiGro,KiWig}, and
Dieperink and Van Isacker~\cite{DiIs}, several terms which are plausible
parts of a mass formula compete to improve the fit to the empirical
masses. It seems safe, though, to conclude that a Wigner term in some
form is called for and that a symmetry energy propotional to $T(T+1)$
is compatible with the data. If a factor $T(T+x)$ is assumed, the
tendency is that $x$ is somewhat less than one. The analyses by
Zeldes~\cite{Ze} and Kirson~\cite{KiWig} also indicate that the
deviation from a quadratic $T$ dependence is not confined to the region
of nuclei with approximately equal $N$ and $Z$, such as implied by the
exponential parametrization of Myers and Swiatecki~\cite{MySwWig} and
the theories of Brenner~\etal~\cite{Br*}, Satu\l a~\etal~\cite{Sa*},
Satu\l a and Wyss~\cite{SaWyWig,SaWyCra}, and Civitarese, Reboiro, and
Vogel~\cite{CiReVo}.

\Section {\label{the}Theory}

\Subsection {\label{ham}Hamiltonian}

The Hamiltonian considered may be thought of as that of a spherical or
deformed shell model. It is written
\begin{gather}
  H = \sum_j \ba \epsilon j \hc a j \ba a j
    + \hf \sum_{ j k l m } \ba v { j k lm }
                           \hc a j \hc a k \ba a m \ba a l \,,
  \label{H}
\end{gather}
where $a_j$ annihilates a nucleon in the state $\ke j$ and the
summations runs over an orthonormal set of single-nucleon states
spanning a valence space. The basic single-nucleon states form
quadruples \nb{\ke j = \ke{q \sigma \tau}} with a common energy
\nb{\epsilon_j = \epsilon_q}, where $q \sigma$ is $q$ or $\rv q$, and
$\tau$ is $\n$, denoting a neutron state, or $\p$, denoting a proton
state. These states are related by
$\ke {\rv q \tau} = \rv {\ke {q \tau}}$ and
$\ke {q \p} = t_- \ke {q \n}$, where a bar over a ket denotes time
reversal, and $t_\pm = t_x \pm i t_y$ in terms of the single-nucleon
isospin $\vc t = (t_x,t_y,t_z)$. Note that time reversal is defined so
as to commute with $t_-$, so $t_x$ and $t_z$ are even and $t_y$ odd
under time reversal.

The interaction matrix element $v_{jklm}$ has a pairing and a
particle-hole part, which are treated differently below,
\begin{gather*}
  \ba v { j k l m } = v^\pair_{ j k l m } + v^\ph_{ j k l m } \,.
\end{gather*}
The pairing part is the matrix element of the isobarically invariant
isovector pairing force,
\begin{gather}
  v^\pair_{ j k l m }
    = - G \me j { \vc t } { \cj k } \cdot \me { \cj m } { \vc t } l
  \,, \quad \ke {\cj \jm} = - 2 i t_y \rv { \ke j } \,.
  \label{vpair}
\end{gather}
The components of
\begin{gather*}
  \vc P = ( P_x , P_y , P_z)
        = \rr \sum_{ j k } \me { \cj k } { \vc t } j \ba a k \ba a j
\end{gather*}
are given by
\begin{gather}
\begin{gathered}
  P_+ = P_x + i P_y
      = \rt \sum_q \ba a { \rv q \p } \ba a { \nr q \p }
      = \rt P_\p \,, \\
  P_- = P_x - i P_y
      = - \rt \sum_q \ba a { \rv q \n } \ba a { \nr q \n }
      = - \rt P_\n \,, \\
  P_z = \rr \sum_q
          ( \ba a { \rv q \p } \ba a { \nr q \n }
          + \ba a { \rv q \n } \ba a { \nr q \p } ) \,.
\end{gathered}
  \label{P}
\end{gather}
It follows from $\me j {\vc t} k = - \me {\cj k} {\vc t} {\cj \jm}$,
which is easily verified, that $\vc P$ commutes as an isovector with the
isospin
\begin{gather*}
  \vc T = ( T_x , T_y , T_z )
    = \sum_{ j k } \me j { \vc t } k \hc a j \ba a k \,.
\end{gather*}
Note
\begin{gather}
\begin{gathered}
  \ba T + = \hc T - = T_x + i T_y = \sum_{ q \sigma }
    \hc a { q \sigma \n } \ba a { q \sigma \p } \,, \\
  T_z = \hf \sum_{ q \sigma } \Bigl(
    \hc a { q \sigma \n } \ba a { q \sigma \n }
      - \; \hc a { q \sigma \p } \ba a { q \sigma \p } \Bigr) \,.
\end{gathered}
  \label{T}
\end{gather}
The particle-hole part of $v_{ j k l m }$ is the matrix element of the
separable interaction of Bohr and Mottelson~\cite{BoMo}, given by
\begin{gather}
  v^\ph_{ j k l m } = \kappa \me j {\vc t} l \cdot \me k {\vc t} m
  \label{vph}
\end{gather}
with a coupling constant $\kappa$. I call this interaction the
``symmetry force''.

The Hamiltonian~\eqref{H} commutes with $\vc T$ and
\begin{gather*}
  \hat A_\va = \sum_j \hc a j \ba a j \,.
\end{gather*}
It therefore has a complete orthogonal set of eigenstates which are also
eigenstates of $\hat A_\va$, $\vc T^2$, and $T_z$. I denote the
eigenvalues of these operators by $A_\va$, $T(T+1)$, and $M_T$. The
numbers $N_\tau$ of valence neutrons and protons are then given by
\nb{N_\tau = A_\va / 2 + 2 m_t M_T} with \nb{m_t = \pm 1/2} for
\nb{\tau = \n} and $\p$. Since we are concerned with isobaric multiplets
where these numbers are even for \nb{M_T = T}, it follows that $A_\va$
is even, $T$ is an integer, and $A_\va/2$ and $T$ have equal parities.
The eigenstates of $H$ form degenerate multiplets with
$M_T = T , T - 1 , \dots , - T$.

If $G = 0$, the lowest eigenvalue of $H$ for given $A_\va$ and $T$ is
\begin{gather}
  E = 2 \sum_\tau \sum_{ \epsilon_q < \lambda_\tau } \epsilon_q
      + \hf \kappa \Bigl( T(T+1) - \tfrac 3 4 A_\va \Bigr) \,,
  \label{E/noG}
\end{gather}
where the Fermi level $\lambda_\tau$ is such that $N_\tau/2$ levels
$\epsilon_q$ satisfy \nb{\epsilon_q < \lambda_\tau} with
\nb{N_\tau = A_\va / 2 + 2 m_t T}. The eigenvalue of the symmetry force
in Eq.~\eqref{E/noG} reflects $\sum_{a \ne b} \vc t_a \cdot \vc t_b
= \nb { \vc T^2 - \sum_a \vc t_a ^2 }$ and \nb{\vc t^2 = 3/4}, where
$\vc t_a$ is the isospin of the $a$th nucleon. The energy $E$ given by
Eq.~\eqref{E/noG} is a convex function of $A_\va$ and $T$ and an
increasing function of $T$ for given $A_\va$. These properties of the
lowest eigenvalue of $H$ are likely to persist for \nb{G \ne 0}. Since
the actual ground state energy, reduced for the electrostatic energy, of
a doubly even nucleus with $N = Z$ is not a convex function of $A$, the
Hamiltonian~\eqref{H} therefore does not reproduce the absolute reduced
energies, but it may reproduce their differences for a fixed $A$.

This Hamiltonian is evidently very schematic. Its relevance for a study
of superfluid isorotation is due to the fact that it obeys the
symmetries that are violated by the pair potential of Cooper pairing,
global gauge invariance and isobaric invariance. Another virtue of the
Hamiltonian is its simplicity, which makes its behavior transparent.
Other symmetries of a more realistic nuclear Hamiltonian than the global
gauge invariance and the isobaric invariance are violated by the present
one. Like every shell model Hamiltonian, it is thus not invariant under
translational and Galilean transformations. If the single-nucleon
energies are derived from a deformed potential, as in the calculations
in Sect.~\ref{def} below, it is also not invariant under rotations in
space. If the Hamiltonian would obey these symmetries, one could have
solutions of the Hartree-Bogolyubov problem which would break the
symmetries and thus give rise to Nambu-Goldstone solutions of the RPA
problem additional to those discussed in Sect.~\ref{gol} below. The
translational and Galilean invariances are necessarily broken by the
solution of the Hartree-Bogolyubov problem. Imposing, in particular,
rotational invariance would be important in a study of rotationally
excited states. For the present study, which deals with nuclear ground
states, and where excited states are considered only in so far as they
are isobaric analogs of ground states, rotational invariance may be
assumed to be less important.

\Subsection {\label{rou}Routhian}

The eigenstates of $H$ with the quantum numbers $A_\va$, $T$, and $M_T$
are also eigenstates of
\begin{gather*}
  \hat R = H - \lambda \hat A_\va - \mu T_z \,,
\end{gather*}
where $\lambda$ and $\mu$ are parameters, and the lowest eigenvalues $E$
and $R$ of $H$ and $\hat R$ for given $A_\va$, $T$, and $M_T$ are
related by
\begin{gather*}
  R = E - \lambda A_\va - \mu M_T \,.
\end{gather*}
If $E$ is a convex function of $A_\va$ and $T$ and an increasing
function of $T$ for given $A_\va$, then for any set of $A_\va$ and $T$
there exists a set of $\lambda$ and $\mu$ with \nb{\mu = 0} for
\nb{T = 0} and \nb{\mu > 0} for \nb{T > 0} so that $R$ is minimal for
this set of $A_\va$ and $T$ and \nb{M_T = T}. Hence the lowest
eigenstates of $H$ for given $A_\va$ and $T$ and with \nb{M_T = T}
is also the lowest eigenstate of $\hat R$ for some values of $\lambda$
and \nb{\mu \ge 0}. Because the allowed values of $A_\va$ and $T$ form a
discrete set, $\lambda$ and $\mu$ are not unique functions of these
variables. They can be chosen freely within certain limits. Pashkevich
and I~\cite{NePa} call a quantity analogous to $R$, involving the
angular momentum, a ``Routhian'', a term borrowed from analytical
mechanics.

\Subsection {\label{vac}Quasinucleon vacuum}

I now set out to calculate $R$ approximately by perturbation theory
starting from a vacuum $\ke \Phi$ of Bogolyubov~\cite{BgTra}
quasinucleons. The state $\ke \Phi$ is determined up to a phase by
\begin{gather}
  \ba \alpha j \ke \Phi = 0
  \label{Phi}
\end{gather}
in terms of a complete set of quasinucleon annihilators
\begin{gather*}
  \ba \alpha j
    = \sum_k \bigl(\ba u {jk} \ba a k + \ba v {jk} \hc a k \bigr)
\end{gather*}
obeying
\begin{gather*}
  \bigl\{ \ba \alpha j , \ba \alpha k \bigr\} = 0 \,, \quad
  \bigl\{ \ba \alpha j , \hc \alpha k \bigr\} = \delta_{jk} \,.
\end{gather*}
It will be assumed to minimize the ``Hartree-Bogolyubov'' Routhian
$R_\HB$ given by
\begin{gather}
  R_\HB = E_\HB - \lambda \Av {\hat A_\va} - \mu \Av {T_z} \,,
  \label{RHB} \\ \bk
  E_\HB = \sum_j \ba \epsilon j \Av {\hc aj \ba aj} \bk
      + \; \hf \sum_{ j k l m } \Bigl(
          v^\pair_{ j k l m } \Av {\hc aj \hc ak} \Av {\ba am \ba al}
          + v^\ph_{ j k l m } \Av {\hc aj \ba al} \Av {\hc ak \ba am}
        \Bigr) \bk
    = \av {H_0} - G \ab {\av {\vc P}} ^2
                + \hf \kappa \av {\vc T} ^2 \,,
    \label{EHB} \\ \bk
  H_0 = \sum_j \ba \epsilon j \hc a j \ba a j \,.
  \nonumber
\end{gather}
In these expressions, the expectation values are in the state $\ke
\Phi$. This is my convention from now on unless otherwise stated in the
context. If $R_\HB$ is minimal, $\ke \Phi$ is an eigenstate of
\begin{gather}
  R_0 = \sum_{ \iota \kappa }
            \pdv {R_\HB} {\av { \ba a \iota \ba a \kappa } }
            \ba a \iota \ba a \kappa \bk
      = H_0 - G \bigl( \cc { \av {\vc P} } {} \cdot \vc P
                + \av {\vc P} \cdot \hc {\vc P} {} \bigr)
        + \kappa \av {\vc T} \cdot \vc T \bk
        - \; \lambda \hat A_\va - \mu T_z \,,
  \label{R0}
\end{gather}
where the convention has been introduced that a Greek letter subscript
takes a value $j$ or $j^{-1}$ with \nb{\ba a {j^{-1}} = \hc a j}. (A
similar notation is used in an early article~\cite{NeVo} by Vogel and
me.) The partial derivatives in Eq.~\eqref{R0} refer to the
expressions~\eqref{RHB} and~\eqref{EHB} with all
$\av { \ba a \iota \ba a \kappa }$ considered mutually independent.
Since $\ke \Phi$ is an eigenstate of $R_0$, the annihilators $\alpha_j$
can be chosen as eigenvectors of the linear map
$\alpha \mapsto [ \alpha , R_0 ]$.

If the isovector $\av { \vc T}$ is different from the zero vector, it
can be rotated into an arbitrary direction by an isobaric transformation
of $\ke \Phi$. This does not change the first two terms in the
expression~\eqref{RHB}. For $\mu > 0$, the isovector $\av { \vc T}$
therefore points into the $z$ direction if $R_\HB$ is minimal. For
$\mu = 0$, the value of $R_\HB$ does not depend on the direction of
$\av { \vc T}$, which can therefore be just assumed to point into the
$z$ direction. Generally, then, \nb{\av {T_x} = \av {T_y} = 0}, which
also holds if $\av { \vc T}$ is the zero vector.

The exact ground state of $\hat R$ is an eigenstate of
\nb{\hat N_\tau = \hat A_\va / 2 + 2 m_t T_z} with even eigenvalues and
therefore an eigenstate of $e^{- i \pi \hat N_\tau}$ with the
eigenvalues one. If $\ke \Phi$ is an eigenstate of $e^{- i \pi \hat
N_\tau}$, the eigenvalues should therefore be one as well; that is,
\begin{gather}
  e^{ - i \pi \hat N_\tau } \ke \Phi = \ke \Phi \,.
  \label{con}
\end{gather}
Suppose $R_\HB$ has been minimized with these constraints, and consider
an infinitesimal variation $\ke {\delta \Phi}$ which violates the
constraints. Any quasinucleon vacuum is an eigenstate of
\nb{e^{ - i \pi \hat A_\va }
= e^{ - i \pi \hat N_\n } e^{ - i \pi \hat N_\p } }, which has the
eigenvalues $\pm 1$. By continuity, therefore,
\nb{e^{ - i \pi \hat N_\n } e^{ - i \pi \hat N_\p } \ke { \delta \Phi }
= \ke { \delta \Phi } }. Since $e^{- i \pi \hat N_\tau}$ has the
eigenvalues $\pm 1$, the variation must then satisfy
\nb{e^{ - i \pi \hat N_\tau } \ke { \delta \Phi }
= - \ke { \delta \Phi } }. It is easily verified that $R_\HB$ is
stationary with respect to such a variation, and it will be seen in
Sect.~\ref{gol} that $R_\HB$ is locally minimized by a state which
satifies the constraints~\eqref{con}. I cannot prove that it is also
globally minimized by this state; this issue is discussed a little
further in Sect.~\ref{gol}. Anyway, I impose from now on the
constraints~\eqref{con}. This entails \nb{\av {P_z} = 0}, so
Eq.~\eqref{R0} becomes
\begin{gather*}
  R_0 = H_0 - \sum_\tau
    ( \cc \Delta \tau \ba P \tau + \ba \Delta \tau \hc P \tau ) \\
    + \; ( \kappa \av {T_z} - \mu ) \cdot T_z
    - \lambda \hat A_\va \,, \quad
  \ba \Delta \tau = G \av {\ba P \tau} \,.
\end{gather*}

A transformation
$\ke \Phi \mapsto e^{-i \sum_\tau \phi_\tau \hat N_\tau } \ke \Phi$
with suitable angles $\phi_\tau$ makes \nb{\ba \Delta \tau \ge 0}. This 
leads to an expression for $R_0$ in the form of the single-quasinucleon
Hamiltonian of the theory of Cooper pairing of Bogolyubov~\cite{BgCoo}
and Valatin~\cite{Va},
\begin{gather}
  R_0 = H_0 - \sum_\tau 
    ( \ba \Delta \tau ( \ba P \tau + \hc P \tau )
    + \lambda_\tau \hat N_\tau ) \,, \bk
  \lambda_\tau = \lambda + m_t ( \mu - \kappa \av {T_z} ) \,,
  \label{lm/tau}
\end{gather}
so that the solution of the eigenproblem
\begin{gather*}
  [ \ba \alpha j, R_0 ] = \ba E j \ba \alpha j \,, \quad
  \ba E j > 0 \,,
\end{gather*}
is obtained immediately from this theory,
\begin{gather}
  \ba \alpha {\nr q \tau} = \ba u {\nr q \tau} \ba a {\nr q \tau}
    - \ba v {\nr q \tau} \hc a {\rv q \tau} \,, \quad
  \ba \alpha {\rv q \tau} = \ba u {\nr q \tau} \ba a {\rv q \tau}
    + \ba v {\nr q \tau} \hc a {\nr q \tau} \,,
  \label{alpha} \\
  \ba u {q \tau} = \sqrt { \hf \biggl( 1 + 
    \frac {\epsilon_q - \lambda_\tau} {E_{q \tau}} \biggr) } \,, \quad
  \ba v {q \tau} = \sqrt { \hf \biggl( 1 -
    \frac {\epsilon_q - \lambda_\tau} {E_{q \tau}} \biggr) } \,, 
  \label{uv}\\
  E_{\nr q \tau} = E_{\rv q \tau}
    = \sqrt { (\epsilon_q - \lambda_\tau) ^ 2 + \Delta_\tau^2}\,.
  \label{Eq}
\end{gather}
Because $\hc \alpha j$ is an eigenvector of
$\alpha \mapsto [ \alpha , R_0 ]$ with the eigenvalue $- \ba E j$,
either $\ba \alpha j$ or $\hc \alpha j$ could annihilate $\ke \Phi$. It
is shown, however, in Sect.~\ref{gol} that a quasinucleon vacuum
annihilated by an operator $\hc \alpha j$ cannot locally minimize
$R_\HB$ if it satisfies Eq.~\eqref{con}. The operators~\eqref{alpha}
therefore annihilate $\ke \Phi$, and Eq.~\eqref{con} holds by the
construction of $\ke \Phi$.

The relations \nb{\Delta_\tau = G \av {P_\tau}} are equivalent to
\begin{gather}
  \sum_q \frac 1 {E_{q\tau}} = \frac 2 G \quad \text{or} \quad 
  \Delta_\tau = 0 \,.
  \label{gap}
\end{gather}
Unless otherwise stated in the context, $\Delta_\tau > 0$ will be
assumed. It remains to fix $\lambda_\tau$. The conventional way of the
theory of Cooper pairing, which I shall follow, consists in demanding
\begin{gather}
  \av {\hat N_\tau} = 2 \sum_q v_{q \tau}^2
    = N_\tau = \frac {A_\va} 2 + 2 m_t T \,.
  \label{N/tau}
\end{gather}
This implies
\begin{gather*}
  E = E_\HB + ( R - R_\HB) \,.
\end{gather*}
In other words, corrections to $E_\HB$ can be calculated as corrections
to $R_\HB$. From Eqs.~\eqref{lm/tau} and~\eqref{N/tau} one gets
\begin{gather}
  \lambda_\tau = \lambda + m_t ( \mu - \kappa T ) \,, \bk
  \lambda = (\lambda_\n + \lambda_\p) / 2 \,, \quad
  \mu = \lambda_\n - \lambda_\p + \kappa T \,,
  \label{mu}
\end{gather}
and Eq.~\eqref{EHB} becomes
\begin{gather}
  E_\HB = E_0 + E_\pair + \hf \kappa T^2 \,, \label{EHB/fin} \\ \bk
  E_0 = \av {H_0}
    =2 \sum_{ q \tau } v^2_{q \tau} \ba \epsilon q \,, \label{E0} \\
  E_\pair = - G \ab {\av {\vc P}} ^2
    = - \frac { \Delta_\n ^2 + \Delta_\p ^2 } G \,.
  \label{Epair}
\end{gather}

If and only if \nb{T = 0}, one has
$u_{ q \n } = u_{ q \p }$, $v_{ q \n } = v_{ q \p }$, and
$E_{ q \n } = E_{ q \p }$. Furthermore, $\lambda_\n - \lambda_\p$ and,
hence, $\mu$ are increasing functions of $T$, so \nb{T = 0},
\nb{\lambda_\n = \lambda_\p}, and \nb{\mu = 0} are equivalent. For 
\nb{T = 0}, also $\Delta_\n = \Delta_\p$, so in terms of its Cartesian
components, $\av { \vc P }$ is purely imaginary and points into the $y$
direction. Other states with the same $R_\HB$ are generated in this case
by arbitrary global gauge transformations and isobaric transformations
of $\ke \Phi$. These transformations generally change the complex
argument and the direction of $\av { \vc P }$. Rotating, in particular,
$\av { \vc P }$ into the $z$ direction by an isobaric transformation
yields $\Delta_\n = \Delta_\p = 0$ and
$\Delta_\np = G \av { P_z } \ne 0$. This solution of the
Hartree-Bogolyubov problem for the isobarically invariant isovector
pairing force is mentioned by Engel~\etal~\cite{En*} as alternative to
the former. Both solutions are in fact just two out of an infinity of
equivalent solutions related by global gauge transformations and
isobaric transformations. In general, these solutions violate the
isobarically noninvariant constraints~\eqref{con}.

\Subsection {\label{rpa}RPA}

In my brief articles~\cite{Ne}, I derive the RPA part of the theory from
a boson expansion. This allows some shortcuts by reference to the
literature. Boson expansions are, however, ambiguous due to the
noncommutability of boson field operators. Following
Thouless~\cite{ThRPA}, I base the following derivation on perturbation
theory.

For any Hamiltonian $\cl H$, and any pair of operators $X$ and $Y$, I
define the propagator
\begin{gather}
\begin{gathered}
  \cl G_t ( X , Y  , t , \cl H ) =
    - i \langle T \{ X'(t) Y'(0) \} \rangle \,, \\
  X(t) = e^{ i \cl H t } X e^{ - i \cl H t } , \quad
  X' = X - \langle X \rangle \,,
\end{gathered}
  \label{prop}
\end{gather}
where $T \{ \dots \}$ indicates time ordering and the expectation
values are in the ground state of $\cl H$. It is easily proved by
Thouless's~\cite{ThMbo} method that $\cl G_t( X , Y , t , \cl H )$ is
the sum of all Feynman diagrams, referring to some independent-particle
Hamiltonian, where the vertices representing $X$ are linked to the
vertices representing $Y$ and with no unlinked part. In particular, the
subtraction of $\langle X \rangle$ in the last of the
equations~\eqref{prop} cancels diagrams without an unlinked part where
the vertices representing $X$ are not linked to the vertices
representing $Y$. If $X$ and $Y$ are linear combinations of products of
even numbers of fermion field operators, the Fourier transform of $\cl
G_t( X , Y , t , \cl H )$ is
\begin{gather*}
  \cl G_{\omega} ( X , Y  , \omega , \cl H )
    = \int_{-\infty}^{\infty}
        e^{ i \omega t } \cl G_t( X , Y , t , \cl H ) d t \\
    = \biggl\langle
        X' \frac 1 { \omega - ( \cl H - \cl E - i \eta ) } Y'
        - Y' \frac 1 { \omega + ( \cl H - \cl E - i \eta ) } X'
      \biggr\rangle \,,
\end{gather*}
where $\cl E$ is the lowest eigenvalue of $\cl H$ and \nb{\eta > 0} is
infinitesimal. I consider
propagators
\begin{gather*}
  G ( X  ,Y , \omega ) = \cl G_{\omega} ( X , Y , \omega , \hat R ) \,,
    \\
  G_0 ( X , Y , \omega ) = \cl G_{\omega} ( X , Y , \omega , R_0 ) \,,
\end{gather*}
and Feynman diagrams referring to $R_0$ as the independent-particle
Hamiltonian. In particular,
\begin{gather*}
  G_0 ( \ba \alpha k \ba \alpha j , \hc \alpha l \hc \alpha m , \omega )
    = \frac
      { \delta_{ j l } \delta_{ k m } -  \delta_{ j m } \delta_{ k l } }
      { \omega - ( \ba E j + \ba E k - i \eta ) } \,, \\ 
  G_0 ( \hc \alpha l \hc \alpha m , \ba \alpha k \ba \alpha j , \omega )
    = \frac
      { \delta_{ j l } \delta_{ k m } -  \delta_{ j m } \delta_{ k l } }
      { - \omega - ( \ba E j + \ba E k - i \eta ) } \,, \\ 
  G_0 ( \ba \alpha \iota \ba \alpha \kappa , 
        \ba \alpha \lambda \ba \alpha \mu , \omega ) = 0 \quad
  \text {otherwise.}
\end{gather*}

The propagator
$G ( \ba a \iota \ba a \kappa , \ba a \lambda \ba a \mu , \omega)$ is
approximated by the sum of the diagrams
\begin{gather}
  \includegraphics{G.mps}
  \label{Ggr}
\end{gather}
where a ``lens'' represents
$G_0 ( \ba a \iota \ba a \kappa , \ba a \lambda \ba a \mu , \omega)$,
and a dashed line the interaction matrix element
$\ba w { \iota \kappa , \lambda \mu }$ given by
\begin{gather}
\begin{gathered}
  \ba w { j^{-1} k^{-1} , m l } =  \ba w { m l , j^{-1} k^{-1} } 
    = \hf v^\pair_{ j k l m } \,, \\
  \ba w { j^{-1} l , k^{-1} m } =  v^\ph_{ j k l m } \,, \\
  \ba w { \iota \kappa , \lambda \mu } = 0  \quad \text {otherwise.}
\end{gathered}
  \label{w}
\end{gather}
Due to the separable form of these matrix elements according to
Eqs.~\eqref{vpair} and \eqref{vph}, the diagrams~\eqref{Ggr} involve
free propagators $G_0 ( X , Y , \omega )$ with many coherent terms over
the configurations of two quasinucleons. They therefore give a large
contribution to the exact propagator
$G ( \ba a \iota \ba a \kappa , \ba a \lambda \ba a \mu , \omega)$.
Summation of these diagrams results in the RPA equation
\begin{gather*}
  G ( \ba a \iota \ba a \kappa , \ba a \lambda \ba a \mu , \omega)
   = G_0 ( \ba a \iota \ba a \kappa ,
           \ba a \lambda \ba a \mu , \omega ) \bk
     + \sum_{ \pi \rho \sigma \tau }
       G_0 ( \ba a \iota \ba a \kappa ,
             \ba a \pi \ba a \rho , \omega )
         \ba w { \pi \rho , \sigma \tau }
       G ( \ba a \sigma \ba a \tau ,
           \ba a \lambda \ba a \mu , \omega ) \,.
\end{gather*}
In a basis of a complete set of independent operators
$\ba \alpha k \ba \alpha j$ and their Hermitian conjugates
$\ba \alpha {j^{-1}} \ba \alpha {k^{-1}}$, this equation takes the
matrix form
\begin{gather}
  \ma G (\omega)
    = \ma G_0 (\omega) + \ma G_0 (\omega) \ma V \ma G (\omega)
    = ( \ma G_0 (\omega)^{-1} - \ma V )^{-1} \,,
  \label{G}
\end{gather}
where
\begin{gather}
  \ba {\ma V} { \iota \kappa , \lambda \mu }
  = \sum_{ \pi \rho \sigma \tau }
      \rb { \ba a \pi \ba a \rho } { \iota \kappa }
      \rb { \ba a \sigma \ba a \tau } { \lambda \mu }
      \ba w { \pi \rho , \sigma \tau }
  \label{V-me}
\end{gather}
with
\begin{gather}
  \rb X { \iota \kappa }
    = \Av { \{ \hc \alpha \kappa \hc \alpha \iota , X \} } \,.
  \label{()}
\end{gather}
Eq.~\eqref{G} implies that $\ma G (\omega)$ has poles where
\begin{gather*}
  \ma G_0 (\omega)^{-1} - \ma V
\end{gather*}
is singular.

It is convenient to introduce also block matrices corresponding to a
division of the basis into a first part consisting of the operators
$\ba \alpha k \ba \alpha j$ and a second part consiting of the operators
$\ba \alpha {j^{-1}} \ba \alpha {k^{-1}}$ with a common order of the
pairs $(j , k)$ in both part. Since
\begin{gather*}
  \rb { \ma G_0 (\omega)^{-1} } { j^{-1} k^{-1} , k j }
    = \omega - ( \ba E j + \ba E k - i \eta ) \,, \\
  \rb { \ma G_0 (\omega)^{-1} } { k j , j^{-1} k^{-1} }
    = - \omega - ( \ba E j + \ba E k - i \eta ) \,, \\
  \rb { \ma G_0 (\omega)^{-1} } { \iota \kappa , \lambda \mu } = 0
  \quad \text {otherwise,}
\end{gather*}
the poles of $\ma G (\omega)$ are then the eigenvalues of
\begin{gather}
  \ma R = \tbt 0 1 {-1} 0 ( \ma E + \ma V ) \,,
  \label{R}
\end{gather}
where $\ma 0$ and $\ma 1$ denote the zero and identity matrices and
\begin{gather}
  \ba {\ma E} { j^{-1} k^{-1} , k j }
    = \ba {\ma E} { k j , j^{-1} k^{-1} }
    = \ba E j + \ba E k - i \eta \,,
  \label{eta} \\
  \ba {\ma E} { \iota \kappa , \lambda \mu } = 0 \quad
  \text {otherwise.}
  \nonumber
\end{gather}

The properties of $\ma R$ and
\begin{gather}
  \ma K = \tbt 0 1 1 0 ( \ma E + \ma V ) = \tbt 1 0 0 {-1} \ma R
  \label{K}
\end{gather}
are discussed by Thouless~\cite{ThRPA}. I shall follow his discussion
partially. Because $\ma E + \ma V$ is symmetric,
\begin{gather*}
  \tp {\ma R} {} = (\ma E + \ma V) \tbt 0 {-1} 1 0
  = - \tbt 0 1 {-1} 0  \ma R \tbt 0 {-1} 1 0  \,.
\end{gather*}
Therefore, ${\ma R}$ has an eigenvalue $- \omega$ for every eigenvalue
$\omega$. For \nb{\eta = 0}, $\ma K$ is Hermitian. If $\ma x$ is a right
eigenvector of $\ma R$ with the eigenvalue $\omega$, therefore
\begin{gather}
  \hc {\ma x} {} \tbt 1 0 0 {-1} \ma R
    = \hc {\ma x} {} \ma K \bk
    = \hc {\ma x} {} \hc {\ma K} {}
    = \hc {\ma x} {} \hc {\ma R} {} \tbt 1 0 0 {-1}
    = \cc \omega {} \hc {\ma x} {} \tbt 1 0 0 {-1} \,,
  \label{l-ev}
\end{gather}
so $\hc { \ma x } {} \tbt 1 0 0 {-1}$ is a left eigenvector with the
eigenvalue $\cc \omega {}$. Two right eigenvectors $\ba {\ma x} 1$ and
$\ba {\ma x} 2$ with the eigenvalues $\ba \omega 1$ and $\ba \omega 2$
satisfy
\begin{gather*}
  \cc \omega 1 \hc {\ma x} 1 \tbt 1 0 0 {-1} \ba {\ma x} 2
    = \hc {\ma x} 1 \tbt 1 0 0 {-1} \ma R \ba {\ma x} 2
    = \ba \omega 2 \hc {\ma x} 1 \tbt 1 0 0 {-1} \ba {\ma x} 2\,,
\end{gather*}
so
\begin{gather}
  \hc {\ma x} 1 \tbt 1 0 0 {-1} \ba {\ma x} 2 = 0 \quad \text{if} \quad
    \cc \omega 1 \ne \ba \omega 2 \,.
  \label{ortho}
\end{gather}
If $\ma K$ is also positive definite, the matrix
\begin{gather*}
  \ma K ^{1/2} \ma R \ma K ^{-1/2}
    = \ma K ^{1/2} { \tbt 1 0 0 {-1} } \ma K ^{1/2}
\end{gather*}
is Hermitian, so the eigenvalues of $\ma R$ are real, and its right
eigenvectors span the space of column matrices of their dimension. 
(This simple argument is due to Ring and Schuck~\cite{RiSh}; 
Thouless~\cite{ThRPA} has a more complicated one.) If
$\ma x$ is a right eigenvector of $\ma R$ with the eigenvalue $\omega$,
one then gets from Eq.~\eqref{l-ev}
\begin{gather}
  \omega \hc {\ma x} {} \tbt 1 0 0 {-1} \ba {\ma x} {}
    = \hc {\ma x} {} \ma K \ba {\ma x} {} > 0 \,,
  \label{nm}
\end{gather}
so $\omega$ cannot be zero. When the imaginary term in Eq.~\eqref{eta}
is included, $\ma R$ acquires the additional term
\begin{gather*}
  \delta \ma R = -i \eta \tbt 1 0 0 {-1} \,.
\end{gather*}
It follows from Eq.~\eqref{l-ev} that the resulting change
$\delta \omega$ of $\omega$ is given by
\begin{gather*}
  \delta \omega \, \hc {\ma x} {} \tbt 1 0 0 {-1} \ba {\ma x} {}
    = \hc {\ma x} {} \tbt 1 0 0 {-1} \delta \ma R \, \ba {\ma x} {}
    = -i \eta \, \ab {\ma x} ^2 \,.
\end{gather*}
A comparison with Eq.~\eqref{nm} then shows that the positive
eigenvalues of $\ma R$ move into the lower imaginary halfplane.

It will be seen in Sect.~\ref{dnt} that $\ma K$ is, actually, only
positive semidefinite for \nb{\eta = 0}. To escape the complications
arising from the eigenvalues zero of $\ma K$ in this limit, it is
understood in the next section that an infinitesimal positive definite
term has been added to $\ma K$.

\Subsection {\label{cor}Correction $R - R_\HB$}

The correction $R - R_\HB$ can be divided into two parts,
\begin{gather*}
  R - R_\HB = \bigl( \Av {\hat R} - R_\HB \bigr)
              + \bigl( R - \Av {\hat R} \bigr) \,.
\end{gather*}
A general result derived by Goldstone~\cite{Gs} implies that the second
bracket in this expression is the sum of all linked Feynman diagrams
without external lines and with at least two interaction lines. The
first bracket is the part of the expectation value of the two-nucleon
interaction in the quasinucleon vacuum that is not included in $R_\HB$.
It is given by
\begin{gather*}
  \Av {\hat R} - R_\HB \\
    = \hf \sum_{ j k l m } \biggl(
      v^\pair_{ j k l m } \Bigl(
          \av { \hc a j \hc  ak \ba a m \ba a l }
        - \av { \hc a j \hc a k} \av { \ba a m \ba a l }
        \Bigr) \\
      + \; v^\ph_{ j k l m } \Bigl(
          \av { \hc a j \hc a k \ba a m \ba a l }
        - \av { \hc a j \ba a l } \av { \hc a k \ba a m }
        \Bigr) \biggr) \\
    = \hf \sum_{ j k l m } \biggl\langle
        \hf v^\pair_{ j k l m } \Bigl( 
            \bigl\{ \hc a j \hc a k - \av {\hc a j \hc a k} ,
                    \ba a m \ba a l - \av {\ba a m \ba a l} \bigr\} \\
          + \; \bigl[ \hc a j \hc a k , \ba a m \ba a l \bigr]
          \Bigr) \\
      + \; v^\ph_{ j k l m } \Bigl(
          \bigl( \hc a j \ba a l - \av { \hc a j \ba a l } \bigr)
          \bigl( \hc a k \ba a m - \av { \hc a k \ba a m } \bigr) \\
        + \; \hc a j \bigl[ \hc a k \ba a m , \ba a l \bigr]
        \Bigr) \biggr\rangle \,.
\end{gather*}
It will be seen that the part
\begin{gather}
  \hf \sum_{ j k l m } \Bigl\langle
      \hf v^\pair_{ j k l m }
      \bigl\{ \hc a j \hc a k - \av { \hc a j \hc a k } ,
         \ba a m \ba a l - \av { \ba a m \ba a l } \bigr\} \bk
    + \; v^\ph_{ j k l m }
      \bigl( \hc a j \ba a l - \av { \hc a j \ba a l } \bigr)
      \bigl( \hc a k \ba a m - \av { \hc a k \ba a m } \bigr)
    \Bigr\rangle
  \label{oy}
\end{gather}
of these terms can be combined with terms in $R - \Av {\hat R}$. The
remainder $c$ can be calculated from Eqs.~\eqref{vpair}, \eqref{vph},
and~\eqref{N/tau},
\begin{gather}
  c = \hf \sum_{ j k l m } \Bigl\langle
    \hf v^\pair_{ j k l m }
      \bigl[ \hc a j \hc a k , \ba a m \ba a l \bigr]
     + \; v^\ph_{ j k l m } \hc a j
       \bigl[ \hc a k \ba a m , \ba a l \bigr]
    \Bigr\rangle \bk
  = \tfrac 3 4 \Bigl( G ( 2 d - A_\va )
    - \hf \kappa A_\va \Bigr) \,,
  \label{c}
\end{gather}
where $4 d$ is the dimension of the valence space. The factor 3/4 is
just $\vc t^2$. It is seen that $c$ does not depend on $T$ and thus does
not contribute to the symmetry energy. The pairing-force part of $c$
vanishes when the valence space is halfway filled. The symmetry-force
part is
$\hf \kappa \big( \sum_{a \ne b} \vc t_a \cdot \vc t_b - \vc T^2 \big)$;
compare the remark after eq.~\eqref{E/noG}.

The expression~\eqref{oy} is equal to the expression~\eqref{term} below
with $n = 1$, which is the first diagram in the series
\begin{gather}
  \includegraphics{R.mps}
  \label{Rgr}
\end{gather}
I approximate $R - R_\HB - c$ by the sum $E_\RPA$ of this series. This
is expected to be a good approximation for the reason mentioned in
connection with the diagrams~\eqref{Ggr}. The $n$th diagram equals
\begin{gather}
  \frac i {2n} \int _{- \infty} ^{\infty}
      \tr \bigl( \ma V \ma G_0 (\omega) \bigr)^n 
    \frac {d \omega} {2 \pi} \,,
  \label{term}
\end{gather}
where the denominator $2 n$ appears because the diagram has $2 n$
equivalent vertices. Hence
\begin{gather}
  E_\RPA = - \frac i 2 \int _{- \infty} ^{\infty} \Biggl( 
      - \sum _{n = 1} ^{\infty}
        \frac 1 n \, \tr \bigl( \ma V \ma G_0 (\omega) \bigr) ^n
    \Biggr) \frac {d \omega} {2 \pi} \,.
  \label{ERPA/int}
\end{gather}
When the positive eigenvalues of $\ma R$ for $\eta = 0$ are denoted by
$\omega_n$, and the set of pairs $( j , k )$ of subscripts of
the basic operators $\alpha_k \alpha_j$ by $\cl S$, the integrand in
Eq.~\eqref{ERPA/int} can be expressed as follows.
\begin{gather*}
   - \sum _{n = 1} ^{\infty}
       \frac 1 n \, \tr \bigl( \ma V \ma G_0 (\omega) \bigr) ^n \\
   = \tr \log \bigl( 1 - \ma V \ma G_0 (\omega) \bigr)
   = \log \det \bigl( 1 - \ma V \ma G_0 (\omega) \bigr) \\
   = \log \det
     \bigl( ( \ma G_0 (\omega) ^{-1} - \ma V )
            \ma G_0 (\omega) \bigr)  \\
   = \log \det \biggl(
      ( \omega - \ma R )
      \Bigl( \omega - \tbt 0 1 {-1} 0 \ma E \Bigr) ^{-1} \biggr) \\
   = \log \Biggl( \prod_n
       \bigl( \omega - ( \ba \omega n -i \eta ) \bigr)
       \bigl( \omega + ( \ba \omega n -i \eta ) \bigr) \\
     \Bigg/ \hspace{-1.5ex} \prod_{ ( j , k ) \in \cl S }
       \bigl( \omega - ( \ba E j + \ba E k - i \eta ) \bigr)
       \bigl( \omega + ( \ba E j + \ba E k - i \eta ) \bigr)
     \Biggr)  \\
   = \sum_n  \Bigl(
         \log \bigl( \omega - ( \ba \omega n -i \eta ) \bigr)
         + \log\bigl( \omega + ( \ba \omega n -i \eta ) \bigr)
       \Bigr) \\
     - \; \sum_{ ( j , k ) \in \cl S } \Bigl(
         \log \bigl( \omega - ( \ba E j + \ba E k - i \eta ) \bigr) \\
         + \; \log \bigl( \omega + ( \ba E j + \ba E k - i \eta ) \bigr)
       \Bigr) \,.
\end{gather*}
Since this is proportional to $\omega ^{-2}$ for large $\omega$, the
integration path can be extended with an infinite semicircle in the
lower imaginary halfplane. When this semicircle is deformed into a
linear path running backward below the real axis at the distance
$2 \eta$, the integral collects contributions only from the
discontinuity of $\log z$ for \nb{z < 0} in the terms in the integrand
which have this cut in the lower imaginary halfplane. The result is
\begin{gather}
  E_\RPA = \hf \Biggl( \sum_{n} \ba \omega n
    - \sum_{ ( j , k ) \in \cl S } ( \ba E j + \ba E k ) \Biggr) \,.
  \label{ERPA/fin}
\end{gather}
This expression has a very simple interpretation if $\ba \omega n$ and
\nb{\ba E j + \ba E k} are conceived as frequencies of harmonic
oscillators. Then, $E_\RPA$ is just the change in total oscillator zero
point energy induced by the interaction of the quasinucleons.

Truncating the expansion in Feynman diagrams to the sum of the diagrams
where a pair of quasinucleons created together is also annihilated
together is equivalent to treating such a pair as a single boson.
Therefore, the result~\eqref{ERPA/fin} is the same as obtained in the
quasiboson approximation. To my knowledge, the derivation above does not
appear in the literature. A somewhat different derivation by contour
integration is due to Shimizu~\etal~\cite{Sh*}. In the quasiboson
approximation, Marshalek~\cite{Ms} derives an expression for the RPA
correction to a Hartree-Bogolyubov energy which corresponds to the
present $c + E_\RPA$ when the difference between the Routhians
considered is taken into account.

\Subsection {\label{mat}Matrices $\ma E$ and $\ma V$}

From now on, $\eta$ is set to zero without further notice, so $\ma E$ is
real. For any operator $X$, I denote by the corresponding sans-serif
symbol $\ma X$ the column matrix of the brackets
$\rb X { \iota \kappa }$ defined by Eq.~\eqref{()}. From
Eqs.~\eqref{vpair}, \eqref{vph}, \eqref{w}, and~\eqref{V-me} then
follows
\begin{gather}
  \ma V
  = - G \Bigl( \tbt 0 1 1 0 \cc { \vc {\ma P} } {}
               \cdot \tp { \vc {\ma P} } {}
             + \ba { \vc {\ma P} } {}
               \cdot \hc { \vc {\ma P} } {} \tbt 0 1 1 0 \Bigr) \bk
    + \; \kappa \vc {\ma T}
                \cdot \tp { \vc {\ma T} } {} \bk
  = - G \Biggl( \sum_\tau \Bigl(
              \tbt 0 1 1 0 \cc { \ma P } \tau \tp { \ma P } \tau
            + \ba { \ma P } \tau \hc { \ma P } \tau \tbt 0 1 1 0
          \Bigr) \bk
          + \; \tbt 0 1 1 0 \cc {\ma P} \sfz \tp {\ma P} \sfz
          + \ba {\ma P} \sfz \hc {\ma P} \sfz \tbt 0 1 1 0
        \Biggr) \bk
     + \kappa \biggl(
                \hf \Bigl( \ba {\ma T} + \tp {\ma T} -
                         + \ba {\ma T} - \tp {\ma T} + \Bigr)
                + \ba {\ma T} \sfz \tp {\ma T} \sfz
              \biggr) \,.
  \label{V}
\end{gather}
It is seen from Eqs.~\eqref{P}, \eqref{T}, \eqref{alpha},
\eqref{uv}, and~\eqref{()} that $\ba { \ma P } \tau$,
$\ba {\ma P} \sfz$, $\ba {\ma T} \pm$, and $\ba {\ma T} \sfz$ are real.
Furthermore, if $\ma X$ is any of these matrices, then, with
\nb{j = q \sigma \tau} and \nb{k = q' \sigma' \tau'}, its elements
$\rb X { k j }$ and $\rb X { j ^{-1} k ^{-1} }$ are nonzero only for
\nb{q' = q} and \nb{\sigma' \ne \sigma}. Therefore,
only within the subspace spanned by the corresponding basic operators do
the eigenvalues of $\ma R$ differ from those of $\tbt 0 1 {-1} 0 \ma E$
and thus contribute to the difference in Eq.~\eqref{ERPA/fin}. Within
this subspace, one can choose \nb{j = \nr q \tau} and \nb{k = \rv q
\tau'}. I call the space spanned by such basic operators with \nb{\tau =
\tau'} the $\tau \tau$ space, that spanned by such basic operators with
\nb{\tau \ne \tau'} the \emph{np} space, and the direct product of the
\emph{nn} and \emph{pp} spaces the \emph{nn}+\emph{pp} space. Since for
any of the aforementioned matrices $\ma X$, the corresponding operator
$X$ is time reversal even, it follows from Eq.~\eqref{()} that
$\rb { X } { \rv q {\p} \nr q \n } = \rb { X } { \rv q {\n} \nr q \p }$
and $\rb { X } { ( \nr q \n ) ^{-1} ( \rv q {\p} ) ^{-1} }
= \rb { X } { ( \nr q \p ) ^{-1} ( \rv q {\n} ) ^{-1} }$. Therefore, if
the basic operators of the \emph{np} space are replaced with their
linear combinations
\begin{gather}
\begin{gathered}
  \ba { \bigl[ \hc \alpha {} \hc \alpha {} \bigr] } { q \rv q \pm }
    = \rr \Bigl( \hc \alpha { \nr q \n } \hc \alpha { \rv q \p }
      \pm \hc \alpha { \nr q \p } \hc \alpha { \rv q \n } \Bigr) \,, \\
  \ba {\bigl[ \alpha \alpha \bigr] } { \rv q q \pm }
    = \rr \Bigl( \ba \alpha { \rv q \p } \ba \alpha { \nr q \n }
      \pm \ba \alpha { \rv q \n } \ba \alpha { \nr q \p } \Bigr) \,,
\end{gathered}
  \label{qq}
\end{gather}
only the corresponding brackets
\begin{gather*}
\rb { X } { \rv q q \pm }
    = \rr \Bigl( \rb { X } { \rv q \p q \n }
      \pm \rb { X } { \rv q \n q \p } \Bigr) \,, \\
\rb { X } { q ^{-1} \rv q ^{-1} \pm }
    = \rr \Bigl( \rb { X } { ( q \n ) ^{-1} ( \rv q \p ) ^{-1} }
      \pm \rb { X } { ( q \p ) ^{-1} ( \rv q \n ) ^{-1} } \Bigr)
\end{gather*}
with the subscript $+$ are different from zero. The only nonzero
elements of $\ma E$ in the basis of the operators~\eqref{qq} are
\begin{gather}
  \ma E_{ q ^{-1} \rv q ^{-1} \pm \, , \, \rv q q \pm }
    = \ma E_{ \rv q q \pm \, , \, q ^{-1} \rv q ^{-1} \pm }
    = \ba E { q \n } + \ba E { q \p } \,.
  \label{Enp}
\end{gather}
For the calculation of $E_\RPA$ one therefore needs to keep only
the part of the \emph{np} space spanned by the operators with the
subscript $+$. From now on, I call this subspace the \emph{np} space.

Eqs.~\eqref{P}, \eqref{T}, \eqref{alpha}, and~\eqref{()} give
\begin{gather}
\begin{gathered}
  \rb { \ba P \tau } { \rv q \tau \nr q \tau }
    = \AV { \ba P \tau
            \hc \alpha { \nr q \tau } \hc \alpha { \rv q \tau } }
    = \sq u { q \tau } \,, \\
  \rb { \ba P \tau } { ( \nr q \tau ) ^{-1} ( \rv q \tau ) ^{-1} }
    = \AV { \ba \alpha { \rv q \tau } \ba \alpha { \nr q \tau }
            \ba P \tau }
    = - \sq v { q \tau } \,, \\
  \rb { \ba P z } { \rv q \nr q + }
    = \AV { \ba P z
            \ba { \bigl[ \hc \alpha {} \hc \alpha {} \bigr] }
                { q \rv q + } }
    = \ba u { q \n } \ba u { q \p } \,, \\
  \rb { \ba P z } { \nr q ^{-1} \rv q ^{-1} + }
    = \AV { \ba { \bigl[ \alpha \alpha \bigr] } { \rv q q + }
            \ba P z }
    = - \ba v { q \n } \ba v { q \p } \,, \\
  \rb { \ba T + } { \rv q \nr q + }
    = \rb { \ba T - } { \nr q ^{-1} \rv q ^{-1} + }
    = \AV { \ba T +
            \ba { \bigl[ \hc \alpha {} \hc \alpha {} \bigr] }
                { q \rv q + } } \\
    = \rt \, \ba v { q \n } \ba u { q \p } \,, \\
  \rb { \ba T - } { \rv q \nr q + }
    = \rb { \ba T + } { \nr q ^{-1} \rv q ^{-1} + }
    = \AV { \ba T -
            \ba { \bigl[ \hc \alpha {} \hc \alpha {} \bigr] }
                { q \rv q + } } \\
    = \rt \, \ba u { q \n } \ba v { q \p } \,, \\
  \rb { \ba T z } { \rv q \tau \nr q \tau }
    = \rb { \ba T z } { ( \nr q \tau ) ^{-1} ( \rv q \tau ) ^{-1} }
    = \AV { \ba T z
            \hc \alpha { \nr q \tau } \hc \alpha { \rv q \tau } } \\
    = 2 m_t \ba u { q \tau } \ba v { q \tau } \,, \\
  \rb { \hat A_\va } { \rv q \tau \nr q \tau }
    = \rb { \hat A_\va } { ( \nr q \tau ) ^{-1} ( \rv q \tau ) ^{-1} }
    = \AV { \hat A_\va
            \hc \alpha { \nr q \tau } \hc \alpha { \rv q \tau } } \\
    = 2 \ba u { q \tau } \ba v { q \tau } \,, \\
  \rb X a  = 0 \quad \text{otherwise,}
\end{gathered}
  \label{()s}
\end{gather}
where $a$ is $\rv q \tau \nr q \tau$,
$( \nr q \tau ) ^{-1} ( \rv q \tau ) ^{-1}$, $\rv q \nr q +$, or
$\nr q ^{-1} \rv q ^{-1} +$. The elements of $\ma V$ connecting the
\emph{nn}+\emph{pp} and \emph{np} spaces are seen to vanish. The part of
$\ma V$ acting in the \emph{nn}+\emph{pp} space is given by the terms in
the sum~\eqref{V} with $\ba {\ma P} \tau$, $\hc {\ma P} \tau$, and
$\ba {\ma T} z$, and the part acting in the \emph{np} space by the terms
with $\ba {\ma P} z$, $\hc {\ma P} z$, and $\ba {\ma T} \pm$. The
nonzero elements of $\ma E$ in the \emph{nn}+\emph{pp} space are
\begin{gather}
  \ma E_{ ( \nr q \tau ) ^{-1} ( \rv q \tau ) ^{-1} \, , \, 
            \rv q \tau \nr q \tau }
    = \ma E_{ \rv q \tau \nr q \tau \, , \,
            ( \nr q \tau ) ^{-1} ( \rv q \tau ) ^{-1} }
    = 2 \ba E { q \tau } \,.
  \label{Enn/pp}
\end{gather}
In the \emph{np} space, they are given by Eq.~\eqref{Enp}. Note that for
$\kappa = 0$, the elements of $\ma V$ connecting the \emph{nn} and
\emph{pp} spaces vanish. When also $\mu = 0$, the \emph{nn}, \emph{pp},
and \emph{np} parts of $\ma E$ or $\ma V$ are identical.

\Subsection {\label{dnt}Definiteness of $\ma K$}

Since the matrices $\ma E$ and $\ma V$ are real, the symmetries
mentioned in Sect.~\ref{rpa} imply
\begin{gather}
  \ma E + \ma V = \tbt B A A B \,,
  \label{EpV}
\end{gather}
where $\ma A$ and $\ma B$ are symmetric. Hence follow
\begin{gather*}
  \ma K = \tbt A B B A \,, \quad
  \ma R = \tbt A B {-B} {-A} \,,
\end{gather*}
whence
\begin{gather}
  \ma K' = \hf \tbt 1 1 1 {-1} \ma K \tbt 1 1 1 {-1}
    = \tbt { A + B } 0 0 { A - B } \,, \label{K'} \\
  \ma R' = \hf \tbt 1 1 1 {-1} \ma R \tbt 1 1 1 {-1}
    = \tbt 0 { A - B } { A + B } 0 \,.
  \nonumber
\end{gather}

I shall prove that $\ma K$ is positive semidefinite. From
\nb{\ba T z = \hc T z} and \nb{\ba T \pm = \hc T \mp} follow
\nb{\ba {\ma T} \sfz = \tbt 0 1 1 0 \ba {\ma T} \sfz} and
\nb{\ba {\ma T} \pm = \tbt 0 1 1 0 \ba {\ma T} \mp}. Since $\kappa$ is
positive, the symmetry force is seen from Eqs.~\eqref{K} and~\eqref{V}
to contribute, then, a positive semidefinite term to $\ma K$. It is
therefore sufficient to consider the case \nb{\kappa = 0}. From the
remark at the end of Sect.~\ref{mat}, it follows that, in this case, the
\emph{nn} and \emph{pp} spaces can be treated separately. Ginoccio and
Wesener~\cite{GiWe} prove that $\ma K$ is positive semidefinite in these
spaces for \nb{\kappa = 0}. The present proof is somewhat different from
theirs and extended to the \emph{np} space. It also gives the dimension
of $\ma K$'s kernel.

With the $\tau \tau'$ parts of $\ma A $ and $\ma B$ denoted by
${ \ma a }_{ \tau \tau' }$ and ${ \ma b }_{ \tau \tau' }$, it is
according to Eq.~\eqref{K'} sufficients to prove that the matrices
${ \ma a }_{ \tau \tau' } \pm { \ma b }_{ \tau \tau' }$ are positive
semidefinite for \nb{\kappa = 0}. Eqs.~\eqref{V}, \eqref{Enp}, 
\eqref{()s}, \eqref{Enn/pp}, and~\eqref{EpV} give
\begin{gather}
  \ba { \ma a } { \tau \tau' } \pm \ba { \ma b } { \tau \tau' }
   = \ba {\ma e } { \tau \tau' }
     - G \ba { \ma p } { \tau \tau' \mp }
         \tp { \ma p } { \tau \tau' \mp } \,,
  \label{apmb}
\end{gather}
where ${ \ma e }_{ \tau \tau' }$ is the diagonal matrix with the 
diagonal elements \nb{E_{ q \tau } + E_{ q \tau' }} and
${ \ma p }_{ \tau \tau' \pm }$ the column matrix with the elements
\begin{gather}
  p_{ q \tau \tau' \pm } =
    u_{ q \tau } u_{ q \tau' } \pm v_{ q \tau } v_{ q \tau' } \,.
  \label{p}
\end{gather}
For any column matrix $\ma x$, thus
\begin{gather*}
  \hc { \ma x } {} ( \ba {\ma a } { \tau \tau' }
                 \pm \ba {\ma b } { \tau \tau' } ) \ma x
    = \hc { \ma x } {} \bigl(
        \ba {\ma e } { \tau \tau' } - G \ba { \ma p } { \tau \tau' \mp }
                                        \tp { \ma p } { \tau \tau' \mp }
                       \bigr) \ma x \\
    = \hc { \ma x } {} \ba { \ma e } { \tau \tau' } \ma x
      - G \bigl| \tp { \ma p } { \tau \tau' \mp } \ma x \bigr| ^2
    \ge \Bigl( 1 - G \tp { \ma p } { \tau \tau' \mp }
                     \iv { \ma e } { \tau \tau' }
                     \ba { \ma p } { \tau \tau' \mp }
        \Bigr) \hc { \ma x } {} \ba { \ma e } { \tau \tau' } \ma x \\
    = \bigl( 1 - G f_{ \tau \tau' \mp }
         \bigr) \hc { \ma x } {} \ba { \ma e } { \tau \tau' } \ma x
\end{gather*}
with
\begin{gather*}
  f_{ \tau \tau' \pm }
    = \tp { \ma p } { \tau \tau' \pm }
           \iv { \ma e } { \tau \tau' }
           \ba { \ma p } { \tau \tau' \pm }
    = \sum_q \frac { \sq p { q \tau \tau' \pm } }
                   { E _{ q \tau } + E _{ q \tau' } } \,,
\end{gather*}
where equality holds if and only if $\ma x$ is proportional to
$\ma e _{ \tau \tau' } ^{-1} \ba { \ma p } { \tau \tau' \mp }$.
Eqs.~\eqref{gap} and~\eqref{p} give
\begin{gather*}
  f_{ \tau \tau' \pm }
    \le \sum_q \frac 1 { E _{ q \tau } + E _{ q \tau' } } \\
    \le \tfrac 1 4 \sum_q \biggr( 
       \frac 1 { E _{ q \tau } } + \frac 1 { E _{ q \tau' } } \biggr)
    = \frac 1 G \,,
\end{gather*}
where, as it follows from the remark at the end of Sect.~\ref{vac}, both
equalities hold if and only if the sign is $+$ and either
\nb{\tau = \tau'} or \nb{\mu = 0}. Hence, the matrices
${\ma a }_{ \tau \tau' } \pm {\ma b }_{ \tau \tau' }$ are positive
semidefinite. Only a matrix
${\ma a }_{ \tau \tau' } - {\ma b }_{ \tau \tau' }$ can have an
eigenvalue zero, which it has if either \nb{\tau = \tau'} or
\nb{\mu = 0}. The multiplicity of these eigenvalues is one. They are
discussed from another point of view in Sect.~\ref{gol}.

Let the symmetry force be included again. Since, as it has now been
shown, $\ma A + \ma B$ is positive definite, one can apply to $\ma R$
one more similarity transformation,
\begin{gather*}
  \ma R'' = \tbt { ( A + B ) ^\mhf } 0 0 { ( A + B ) ^\mmhf } \ma R'
            \tbt { ( A + B ) ^\mmhf } 0 0 { ( A + B ) ^\mhf } \\
    = \tbt 0 { ( A + B ) ^\mhf ( A - B ) ( A + B ) ^\mhf } 1 0
    =  \tbt 0 M 1 0 \,.
\end{gather*}
The matrix $\ma M$ is real, symmetric, and positive semidefinite. It
therefore has only nonnegative eigenvalues $\omega ^2$, and the
corresponding eigenvectors $\ma x$ span the space of column matrices
of their dimension. If \nb{\omega ^2 > 0}, then
\begin{gather*}
  \tbo { \omega x } x
\end{gather*}
are eigenvectors of $\ma {R''}$ with the eigenvalues $\omega$. Note
that $\omega$ can have both signs, so two eigenvalues of $\ma R''$ with
opposite signs correspond to each such eigenvalue of $\ma M$. If
\nb{\omega ^2 = 0}, one has
\begin{gather*}
  \ma R'' \tbo 0 x = 0 \,, \quad \ma R'' \tbo x 0 = \tbo 0 x \,.
\end{gather*}
The column matrix in the first of these equations is an eigenvector of
$\ma R''$ with the eigenvalue zero, whereas the first one in the second
equation is not an eigenvector of $\ma R''$. Following
Thouless~\cite{ThRPA,ThSta} and Marshalek and Wesener~\cite{MsWe}, one
can interpret the linear combinations of two-quasinucleon annihilators
and creators corresponding to these column matrices as proportional to
leading terms in expansions of a conserved momentum and its conjugate
coordinate. Together, all the pairs of column matrices mentioned above
span the space of column matrices of their dimension.

It follows from Eq.~\eqref{K} that $\ma K$ and $\ma R$ have a common 
right kernel, whose dimension I denote by $d_\tk$. For \nb{\kappa=0},
the preceding discussion implies \nb{d_\tk = 2} if \nb{\mu > 0} and
\nb{d_\tk = 3} if \nb{\mu = 0}. Since $d_\tk$ is also the dimension of
the right kernel of $\ma M$ and
\begin{gather*}
  \det ( \omega -\ma R ) = \det ( \omega -\ma R'' )
    = \det \bigl( \omega ^2 - \ma M \bigr) \,,
\end{gather*}
the characteristic root zero of $\ma R$ has the multiplicity $2 d_\tk$.
When $\ma K$ is made positive definite by an infinitesimal perturbation,
as it was assumed to have been done for the purpose of the derivation in
Sect.~\ref{cor}, the eigenvalue zero of $\ma R$ then splits into $d_\tk$
pairs of nonzero infinitesimal real eigenvalues with opposite signs.
Being infinitesimal, these eigenvalues do not contribute to $E_\RPA$, 
which can therefore be calculated from the positive eigenvalues of the
original singular matrix $\ma R$. These are the square roots of the
positive eigenvalues of $\ma M$. The eigenvalues of $\ma M$ are
calculated most efficiently from
\begin{gather}
  \ma M' = ( \ma A + \ma B ) ^{ 1/2 } \ma M ( \ma A + \ma B ) ^{ -1/2 }
    = ( \ma A + \ma B ) ( \ma A - \ma B ) \,.
  \label{M'}
\end{gather}
That the dimension of the matrix to be diagonalized can be reduced in 
this way when $\ma R$ is real is pointed out by Ring and 
Schuck~\cite{RiSh} in the case when $\ma K$ is positive definite.

\Subsection {\label{gol}Nambu-Goldstone solutions}

It follows from a theorem proved by Thouless~\cite{ThSta} that a general
Bogolyubov quasinucleon vacuum in the vicinity of $\ke \Phi$ is
obtained from $\ke \Phi$ by a transformation of the form
\begin{gather}
  \ke \Phi \mapsto \mathcal N \exp \left( \sum_{ ( j , k ) \in \cl S }
    \ba \zeta { k j } \hc \alpha j \hc \alpha k \right) \ke \Phi \,,
  \label{trf}
\end{gather}
where $\ba \zeta { k j }$ are complex parameters, and $\mathcal N$ is a
normalization factor. Since expectation values in the transformed vacuum
do not depend on its phase, they are functions of $\ba \zeta { k j }$.
Thouless's results in Ref.~\cite{ThSta} imply that $\ma E + \ma V$ is
the Hessian matrix of $R_\HB$ with respect to $\ba \zeta { k j }$ and
$\ba \zeta { j^{-1} k^{-1} } = \cc \zeta { k j }$ for $\ba \zeta { k j }
= 0$ and that, if some symmetry of $E_\HB$ is violated by the
quasinucleon vacuum, special solutions of the eigenproblem of $\ma R$
result from considering variations of $\ke \Phi$ within the symmetry
group. These solutions are analogous to the Nambu-Goldstone~\cite{NaGo}
boson solutions in field theories with a so-called ``spontaneous
symmetry breaking'', that is, a violation of a symmetry by the vacuum.

In the present case, the variations to be considered are
\begin{gather}
  \ke { \delta \Phi } = -i \, \delta \chi X \ke \Phi \,,
  \label{sym-tr}
\end{gather}
where $X$ is any one of the Hermitian operators $\hat N_\tau$, $T_x$ and
$T_y$, and $\delta \chi$ is infinitesimal and real. Because the first
two terms in the expression~\eqref{RHB} are invariant under global gauge
transformations and isobaric transformations,
\begin{gather*}
  \frac {\delta R_\HB} { \delta \chi } =
    - \mu \frac { \delta \av {T_z} } { \delta \chi }
    = i \mu \av Y \,, \quad Y = [ T_z , X ] \,,
\end{gather*}
whence for $\ba \zeta { k j } = 0$, by
$\partial R_\HB / \partial \ba \zeta { \iota \kappa } = 0$, follows
\begin{gather}
  \sum_\lambdamus
    \frac { \partial ^2 R_\HB }
          { \partial \ba \zeta \iotakappa
            \partial \ba \zeta { \lambda \mu } }
    \frac { \delta \ba \zeta { \lambda \mu } }
          { \delta \chi } \bk
  = \sum_\lambdamus
      \biggl(
        \frac { \partial }
              { \partial \ba \zeta \iotakappa }
      \biggl(
        \frac { \partial R_\HB }
              { \partial \ba \zeta { \lambda \mu } }
        \frac { \delta \ba \zeta { \lambda \mu } }
              { \delta \chi }
      \biggr) \bk
    -
      \biggl(
        \frac { \partial }
              { \partial \ba \zeta \iotakappa }
        \frac { \delta \ba \zeta { \lambda \mu } }
              { \delta \chi }
      \biggr)
        \frac { \partial R_\HB }
              { \partial \ba \zeta { \lambda \mu } }
      \biggr) \bk
  = \sum_\lambdamus
      \frac { \partial }
            { \partial \ba \zeta \iotakappa }
       \biggl(
         \frac { \partial R_\HB }
               { \partial \ba \zeta { \lambda \mu } }
         \frac { \delta \ba \zeta { \lambda \mu } }
               { \delta \chi }
       \biggr) \bk
  = \frac { \partial } { \partial \ba \zeta \iotakappa }
    \frac { \delta R_\HB } { \delta \chi }
  = i \mu \frac { \partial \av Y }
                { \partial \ba \zeta \iotakappa }
  = i \mu \rb Y { \iota \kappa } \,.
  \label{gol/der}
\end{gather}
For $\ba \zeta { k j } = 0$, Eqs.~\eqref{trf}
and~\eqref{sym-tr} give
\begin{gather*}
  \delta \ba \zeta { k j }
    = \me { \Phi }  { \ba \alpha k \ba \alpha j } { \delta \Phi }
    = -i \, \delta \chi 
        \av { \ba \alpha k \ba \alpha j  X } \\
    = -i\, \delta \chi \rb X { j ^{-1} k ^{-1} } \,, \\
  \delta \ba \zeta { j ^{-1} k ^{-1} } = \delta \cc \zeta { k j }
    = i \,\delta \chi 
        \av { X  \hc \alpha j \hc \alpha k }
    = i \, \delta \chi \rb X { k j } \,.
\end{gather*}
Eq.~\eqref{gol/der} can therefore be written
\begin{gather*}
  ( \ma E + \ma V ) { \tbt 0 {-i} i 0 } \ma X = i \mu \ma Y \,,
\end{gather*}
or
\begin{gather*}
  \ma R \tbt 0 1 {-1} 0 \ma X = - \mu \tbt 0 1 {-1} 0 \ma Y \,.
\end{gather*}

Explicitly, these relations are
\begin{gather}
\begin{gathered}
  \ma R \tbt 0 1 {-1} 0 \ba { \ma { \hat N } } \tau = 0 \,, \quad
  \ma R \tbt 0 1 {-1} 0 \ba { \ma T } \sfx
    = - i \mu \tbt 0 1 {-1} 0 \ba { \ma T } \sfy \,, \\
  \ma R \tbt 0 1 {-1} 0 \ba { \ma T } \sfy
    = i \mu \tbt 0 1 {-1} 0 \ba { \ma T } \sfx \,.
\end{gathered}
  \label{gol/explc}
\end{gather}
The column matrices $\ba { \ma { \hat N } } \tau$ and $\ba { \ma T }
\mp$ do not vanish because $\ke \Phi$ is not an eigenstate of $\hat
N_\tau$ or $T_\pm$. Therefore
$\tbt 0 1 {-1} 0 \ba { \ma { \hat N } } \tau$ and
$\tbt 0 1 {-1} 0 \ba { \ma T } \mp$ are right eigenvectors of $\ma R$
with the eigenvalues zero and $\pm \mu$. They belong to the $\tau \tau$
and \emph{np} spaces, respectively. For \nb{\mu = 0}, one has
$\ba { \ma T } \sfy = 0$ because $\av { \vc P }$ points into the $y$
direction so that $\ke \Phi$ is an eigenvector of $T_y$. The column 
matrices $\tbt 0 1 {-1} 0 \ba { \ma T } \mp$ then merge into a single
right eigenvector $\tbt 0 1 {-1} 0 \ba { \ma T } \sfx$ of $\ma R$ with
the eigenvalue zero.

For \nb{\kappa = 0}, the colomn matrices in the common right kernel of
$\ma R$ and $\ma K$ found here must be those whose existence was proved
in Sect.~\ref{dnt}. It is now seen that they persist for
\nb{\kappa \ne 0}. It follows from the discussion in Sect.~\ref{dnt} 
that they span the kernel. The derivatives ${ \partial ^2 R_\HB }
/ { \partial \cc \zeta \iotakappa \partial \ba \zeta { \lambda \mu } }$
are for $\ba \zeta { k j } = 0$ the elements of $\ma K$. Since
it has now been proved that $\ma K$ is positive semidefinite and its
entire kernel stems from symmetries of $R_\HB$, it follows that
$\ke \Phi$ minimizes $R_\HB$ locally. For a valence space consisting of
a single $j$-shell, and an energy functional without the symmetry force
and including the complete expectation value of the pairing force,
Camiz, Covello, and Jean~\cite{CaCoJe} prove that a state with
annihilators of the form~\eqref{alpha} minimizes this functional
globally on a certain class of Bogolyubov quasinucleon vacua with
contant $\av {\hat N_\tau }$. Using their method, one can easily prove
that $\ke \Phi$ minimizes $E_\HB$ globally on the same class of
Bogolyubov quasinucleon vacua.

I promised in Sect.~\ref{vac} to prove that a quasinucleon vacuum
annihilated by an operator $\hc \alpha j$ and obeying Eq.~\eqref{con}
cannot minimize $R_\HB$ locally. Assume, to the contrary, that
$\ke \Phi$ is annihilated by an operator $\hc \alpha j$. In order to
satisfy Eq.~\eqref{con}, the number of such operators must be even for
each kind of nucleon, so for one kind of nucleon there must be at least
two such operators, $\hc \alpha {q \tau}$ and $\hc \alpha {q' \tau}$,
say. These operators replace
$\ba \alpha {q \tau}$ and $\ba \alpha {q' \tau}$ in Eq.~\eqref{trf}.
Matrices $\ma E$ and $\ma V$ can be defined as the Hessian matrices of
$\av { R_0 }$ and $R_\HB - \av { R_0 }$ with respect to
$\zeta_{ \iota \kappa }$ for $\zeta _{ k j } = 0$. In
particular,
\begin{gather*}
  \ma E _{ ( q \tau )^{-1} ( q' \tau )^{-1} , q' \tau q \tau }
    = \ma E _{ q' \tau q \tau , (q \tau)^{-1} (q' \tau)^{-1} } \\
    = - \; E _{ q \tau }  - E _{ q' \tau } < 0 \,,
\end{gather*}
while $\ma V$ is still given by an expression in the form of
Eq.~\eqref{V} with real $\ba { \ma P } \tau$, $\ba {\ma P} \sfz$,
$\ba {\ma T} \pm$, and $\ba {\ma T} \sfz$. Now, let a column matrix
$\ma x$ be defined by
\begin{gather*}
  \ma x _{ q' \tau q \tau }
     = -\ma x _{ (q \tau)^{-1} (q' \tau)^{-1} } = 1 \,, \quad
  \ma x _{ \iota \kappa } = 0 \quad \text{otherwise.}
\end{gather*}
Because $T_z$ is Hermitian,
$\tbt 0 1 1 0 \ba {\ma T} z = \ba {\ma T} z$, so
\begin{gather*}
  \tp { \ma T } \pm \ma x = \tp { \ma T } z \ma x = 0 \,.  
\end{gather*}
The contribution to $\tp { \ma x } {} \tbt 0 1 1 0 \ma V \ma x$ from the
symmetry force therefore vanishes. The contribution from the pairing
force is nonpositive. With $\ma K$ given by Eq.~\eqref{K}, then
\begin{gather*}
  \tp { \ma x } {} \ma K \ma x < 0 \,,
\end{gather*}
so $\ma K$ is not positive semidefinite and $\ke \Phi$ does not
minimize $R_\HB$ locally.

The eigenvalues zero of $\ma R$ in the $\tau \tau$ spaces can be
understood to result from the ground state of $\hat R$ being an
eigenstate of $\hat N_\tau$. The similar applies to the eigenvalues zero
in the \emph{np} space for $\mu = 0$. For \nb{\mu > 0}, the eigenvalues 
$\pm \mu$ can be understood to originate in the $M_T = T - 1$ isobaric
analog of the ground state, which has the excitation energy $\mu$ as
an eigenstate of $\hat R$. This interpretation differs from that of
Ginoccio and Wesener~\cite{GiWe}, who, in the quasiboson picture, find
that the vibrational quantum in question increases $T$ by one unit. They
infer this from the commutation relation, which they do not prove,
written in the line after their Eq.~(88b).

In the oscillator interpretation of the expression~\eqref{ERPA/fin}, the
term $\hf \mu$ arises from the quantal fluctuations of the variables
$T_\pm$, which are in the approximation
$[ T_+ , T_- ] = 2 T_z \approx 2 T$ proportional to conjugate
coordinates. In an eigenstate of $\vc T ^2$ and $T_z$, these
fluctuations are determined by the isospin algebra. Assume, in
accordance with the isorotational picture, that intrinsic classical
variables, invariant under isorotation, could be defined so that the
Hamiltonian is a function of $\ab { \vc T }$ and these intrinsic
variables, and let $E_\col ( \ab { \vc T } )$, where ``col'' stands for
``collective'', be the energy of the intrinsic equilibrium for given
$\ab { \vc T }$, where $\vc T$ is a ``classical'' isospin. The term
$\hf \mu$ in Eq.~\eqref{ERPA/fin} then correponds to the following
estimate of the energy due to the quantal fluctuations of $\vc T$.
\begin{gather*}
  E_\col \Bigl( \sqrt { T(T+1) } \Bigr) - E_\col ( T )
    \approx \hf \frac { \Delta E_\col ( T ) } { \Delta T } \\
    \approx \hf \frac { \Delta E } { \Delta T }
    \approx \hf \mu \,,
\end{gather*}
where the differences are taken between the discrete allowed values of
$T$ and the $T$ dependence of the total zero point energy is neglected
in the second approximation. A related interpretation of the a term
involving the angular velocity which emerges analogously in models with
rotational invariance is discussed by Marshalek~\cite{Ms} in a boson
expansion picture.

Since $\ba { \ma T } \sfz = \hf \Bigl(
\ba { \ma { \hat N } } \sfn - \ba { \ma { \hat N } } \sfp \Bigr)$ and
\begin{gather*}
  \tbt 0 1 {-1} 0 \ma {T_\pm}
    = \tbt 0 1 {-1} 0 \tbt 0 1 1 0 \ma {T_\mp}
    = \tbt 1 0 0 {-1} \ma {T_\mp} \,, \\
  \tbt 0 1 {-1} 0 \ma {T_z} = \tbt 0 1{-1} 0 \tbt 0 1 1 0 \ma {T_z} 
    = \tbt 1 0 0 {-1} \ma {T_z} \,,
\end{gather*}
it follows from Eq.~\eqref{ortho} that any other right eigenvector of
$\ma R$ than those mentioned so far is orthogonal to the components of
$\vc {\ma T}$. As seen from Eqs.~\eqref{R} and~\eqref{V}, it then
belongs to the kernel of the symmetry-force term in Eq.~\eqref{V}.
Therefore, $\pm \mu$ are the only eigenvalues of $\ma R$ which depend on
$\kappa$ and the right eigenvectors of $\ma R$ are independent of
$\kappa$. Since, as seen from Eqs.~\eqref{uv}, \eqref{Eq}, \eqref{gap},
and~\eqref{N/tau}, $\lambda_\tau$ and $\Delta_\tau$ are independent of
$\kappa$, it hence follows that it is sufficient to determine the
eigenvalues of $\ma R$ for \nb{\kappa = 0}, in which case the \emph{nn}
and \emph{pp} spaces separate. The symmetry force is then taken into
account by adding the term $\kappa T$ in Eq.~\eqref{mu}.

This is a remarkable result. Its interpretation in the isorotational
picture is that the symmetry force does not influence the intrinsic
excitations. It only contributes to the collective energy. This is
consistent with the fact, following from the remark after
Eq.~\eqref{E/noG}, that the symmetry force differs from
$\hf \kappa \vc T ^2$ only by a constant.

From Eqs.~\eqref{mu}, \eqref{EHB/fin}, \eqref{c}, and~\eqref{ERPA/fin}
follow
\begin{gather}
  E_\RPA = E_{\RPA,\kappa=0} + \hf \kappa T \,, \label{ERPA/kap} \\
  E = E_\HB + c + E_\RPA
    = E_{\kappa=0}
      + \hf \kappa \Bigl( T(T+1) - \tfrac 3 4 A_\va \Bigr) \,.
  \label{E}
\end{gather}
Using Eq.~\eqref{EpV},
$\ba { \ma T } \sfz = \tbt 0 1 1 0 \ba { \ma T } \sfz$, and the fact
that $\tbt 0 1 {-1} 0 \ba { \ma T } \sfz$ belongs to the kernel of
$\ma E + \ma V$, one can easily prove that $\ma M'$ defined by
Eq.~\eqref{M'} is independent of $\kappa$ in the \emph{nn}+\emph{pp}
space. Thus, its elements between the \emph{nn} and \emph{pp} spaces
vanish.

\Subsection {\label{npg}Case $\Delta_\tau = 0$}

Most of the preceding discussion applies for $\Delta_\tau = 0$. The
Fermi level $\lambda_\tau$ can be placed, in this case, anywhere between
such a pair of single-nucleon levels that $N_\tau/2$ levels $\epsilon_q$
satisfy \nb{\epsilon_q < \lambda_\tau}. The pairing energy $E_\pair$
vanishes according to Eq.~\eqref{Epair}, and one gets from
Eqs.~\eqref{EHB/fin} and~\eqref{E0}
\begin{gather}
  E_\HB = 2 \sum_\tau \sum_{ \epsilon_q < \lambda_\tau } \epsilon_q
    + \hf \kappa T^2 \,.
  \label{EHB/noG}
\end{gather}
The column matrices $\ma N_\tau$ vanish, so $\ma R$ has no eigenvalue
zero in the \emph{nn}+\emph{pp} space. The brackets
$\rb { \ba T - } { \rv q \nr q + }$ also vanish, and
$\rb { \ba T - } { \nr q ^{-1} \rv q ^{-1} + }$ is equal to $\sqrt 2$ if
\nb{\lambda_\p < \epsilon_q < \lambda_\n} and zero otherwise. For
$T = 0$, the column matrix $\ba { \ma T } \sfx$ then vanishes because no
level $\epsilon_q$ satisfies \nb{\lambda_\p < \epsilon_q < \lambda_\n},
so $\ma R$ has neither in the \emph{np} space an eigenvalue zero. For
$T > 0$, consider the part of the \emph{np} space spanned by the
operators
$\ba { \bigl[ \hc \alpha {} \hc \alpha {} \bigr] } { q \rv q \pm }$ and
$\ba {\bigl[ \alpha \alpha \bigr] } { \rv q q \pm }$ with
\nb{\lambda_\p < \epsilon_q < \lambda_\n}. In this subspace,
$\tbt 0 1 {-1} 0 \ba { \ma T } -$ has the elements $\sqrt{2}$ in its
upper part and zero in its lower part. From
$\rb { \ba P z } { \rv q \nr q + }
= \rb { \ba P z } { \nr q ^{-1} \rv q ^{-1} + } = 0$ follows
\nb{\ma R = \tbt E 0 0 { -E }} for \nb{\kappa = 0}, where the diagonal
elements of $\ma E$ are $E_{ q \n } + E_{ q \p }
= \nb{ ( \lambda_\n - \epsilon_q ) + ( \epsilon_q - \lambda_\p ) }
= \lambda_\n - \lambda_\p$. Thus $\tbt 0 1 {-1} 0 \ba { \ma T } -$ is
for \nb{\kappa = 0} an eigenvector of $\ma R$ with the eigenvalue
\nb{\lambda_\n - \lambda_\p = \mu}. This is the only positive eigenvalue
of $\ma R$ which depends on $\kappa$. For \nb{\kappa > 0}, it becomes
\nb{\lambda_\n - \lambda_\p + \kappa T = \mu}.

For \nb{G = \kappa = 0}, one has \nb{E_\RPA = 0}. Generally for
\nb{G = 0}, Eq.~\eqref{ERPA/kap} then gives
\begin{gather}
  E_\RPA = \hf \kappa T \,.
  \label{ERPA/noG}
\end{gather}
Eqs.~\eqref{c}, \eqref{E}, \eqref{EHB/noG}, and~\eqref{ERPA/noG} imply
Eq.~\eqref{E/noG}, so the RPA is exact in this case. For $G >0$, the
pairing force gives an additional contribution to $E_\RPA$.

\Subsection {\label{phs}Case of particle-hole symmetry}

Special phemonema occur when the single-nucleon spectrum is symmetric
about some energy and the valence space is halfway filled. For brevity, 
I call this symmetry of the single-nucleon spectrum a
``particle-hole symmetry''. Without loss of generality, the spectrum 
can be assumed, in this case, to be centered at zero and labeled so that
\nb{\epsilon_q = - \epsilon_{ -q } > 0} for \nb{q > 0}. Then,
\nb{\lambda = 0}, \nb{\lambda_\p = - \lambda_\n},
$\Delta_\p = \Delta_\n$, \nb{E_ { q \p } = E_ { (-q) \n }},
\nb{u_{ q \p } = v_{ (-q) \n }}, \nb{v_{ q \p } = u_{ (-q) \n }}, and
$\ma M'$ defined by Eq.~\eqref{M'} has equal eigenvalues in the
\emph{nn} and \emph{pp} spaces.

In the \emph{np} space, the eigenvalues of $\ma M'$ turn out to have the
multiplicity 2 with the exception of the eigenvalue $\mu ^2$ found in
Sect.~\ref{gol} and an eigenvalue
$4 ( \lambda _\n ^2 + \Delta _\n ^2 )$, both of which are for
$\kappa = 0$ less than all $( E_{ q \n } + E_{ q \p } ) ^2$ and less
than all other eigenvalues of $\ma M'$. To see how this happens, first
consider the case \nb{\kappa = 0}. Let
${\ma a }_{ \tau \tau' } \pm {\ma b }_{ \tau \tau' }$ be divided into
blocks corresponding to an ordering of the subscripts $q$ by their signs
with \nb{q > 0} first and a common order of $\ab{q}$ for each
sign. Eqs.~\eqref{apmb} and~\eqref{p} then give
\begin{gather*}
  \hf \tbt 1 1 1 {-1}
      ( { \ma a}_{ \n \p } + { \ma b }_{ \n \p } )
      ( { \ma a}_{ \n \p } - { \ma b }_{ \n \p } )
      \tbt 1 1 1 {-1} \\
  = \tbt e 0  0 { e - \twoG \ba p - \tp p - }
    \tbt { e - \twoG \ba p + \tp p + } 0 0 e \\
  \sim \tbt { \sq e {} - \twoG \sr e {} \ba p + \tp p + \sr e {} } 0
    0 { \sq e {} - \twoG  \sr e {} \ba p - \tp p - \sr e {} } \,,
\end{gather*}
where the similarity is effected by the matrix
$\tbt { \ir e {} } 0 0 { \sr e {} }$. The matrices $\ma e$ and
${ \ma p } _\pm$ are the upper left and upper parts of
$\ma e _{ \n \p }$ and $\ma p _{ \n \p \pm }$. From Eqs.~\eqref{uv},
\eqref{Eq}, and~\eqref{p}, one gets by some algebra
\begin{gather}
  ( E_{ q \n } + E_{ q \p } ) p_{ q \n \p \pm } ^2 \bk
    = \tfrac 1 4
      \biggl( \frac 1 { E_{ q \n } } + \frac 1 { E_{ q \p } } \biggr)
      \bigl( ( E_{ q \n } + E_{ q \p } ) ^2 - \Lambda _\pm \bigr)
  \label{Epsq}
\end{gather}
with
\begin{gather}
  \Lambda _+ = 4 \lambda _\n ^2 \,, \quad
  \Lambda _- = 4 \bigl( \lambda _\n ^2 + \Delta _\n ^2 \bigr) \,.
  \label{Lms}
\end{gather}
It follows from Eq.~\eqref{gap} that the column matrix $\ma s$ with the
dimension of $\ma p _{ \n \p \pm }$ and the elements
\begin{gather}
  s_q = \sqrt{ \hf G \biggl( \frac 1 { E_{ q \n } }
                     + \frac 1 { E_{ q \p } } \biggr) }
  \label{s}
\end{gather}
is a unit vector. From Eq.~\eqref{uv} one can derive
\nb{u_{ q \n } u_{ q \p } > v_{ q \n } v_{ q \p }} for \nb{q > 0}. It
then follows from Eq.~\eqref{p} that the elements of $\ma p _\pm$ are
positive. It follows from Eq.~\eqref{Epsq} that $\Lambda _\pm$ are then
less than all $( E_{ q \n } + E_{ q \p } ) ^2$, so the diagonal matrices
\begin{gather}
  \ma f _\pm = \ma e^2 - \Lambda _\pm
  \label{f}
\end{gather}
are positive definite and the positive eigenvalues of
\begin{gather*}
  \ma g = ( \ma 1 - \ma s \tp { \ma s } {} ) \ma e^2
          ( \ma 1 - \ma s \tp { \ma s } {} )
\end{gather*}
are greater than $\Lambda _\pm$.

From Eqs.~\eqref{Epsq}, \eqref{s}, and~\eqref{f}, one gets
\begin{gather*}
  \ma e ^2 - \twoG \ma e ^{1/2} \ma p _\pm \ma p _\pm ^T \ma e ^{1/2}
  = \ma f _\pm - \ma f _\pm ^{1/2} \ma s \ma s ^T \ma f _\pm ^{1/2}
    + \Lambda _\pm \bk
  \sim \ma f _\pm \bigl( \ma 1 - \ma s \ma s ^T \bigr)
       + \Lambda _\pm = \ma h _\pm \,.
\end{gather*}
It is seen that $\ma s$ is an eigenvector of $\ma h _\pm$ with the
eigenvalue $\Lambda _\pm$. Furthermore, if $\omega ^2 > 0$ is an
eigenvalue of $\ma g$ and $\ma x$ the corresponding eigenvector, then
\begin{gather*}
  \ma x - \frac 1 { \omega ^2 - \Lambda _\pm } 
    \ma s \ma s ^T \ma e ^2 \ma x
\end{gather*}
is an eigenvector of $\ma h _\pm$ with this eigenvalue. The matrices
$\ma h _\pm$ thus have these eigenvalues in common. The eigenvalue
$\Lambda _+ = ( 2 \lambda _\n ) ^2 = \mu ^2$ of $\ma M'$ is just that
discussed in Sect.~\ref{gol}. It is the only eigenvalue of $\ma M'$
which depends on $\kappa$. For \nb{\kappa > 0}, it becomes
\nb{( 2 \lambda _\n + \kappa T ) ^2 = \mu ^2}. The eigenvalues
$\omega ^2$ of $\ma h_\pm$ are seen to satisfy
\begin{gather}
  \ma p _\pm ^T \frac { 2 \ma e } { \ma e ^2 - \omega ^2 } \ma p _\pm
  = \frac  1 G \,.
  \label{se}
\end{gather}
Thus, $\ma M'$ has in the \emph{np} space one twofold degenerate
eigenvalue in each interval between consecutive squares of
\nb{E_{ q \n } + E_{ q \p } , \ q > 0}.

For \nb{T = 0}, the eigenvalues of $\ma M'$ in a $\tau \tau$ space
and their multiplicities are the same as in the \emph{np} space. The
eigenvalues with the multiplicity one are zero and $4 \Delta_\n^2$ in
this case. It follows that, more generally, if the energy levels of the
nucleons with isospin $m_t$ are symmetrically distributed about
$\lambda_\tau$, then the eigenvalues of $\ma M'$ in the $\tau \tau$
space have the multiplicity two with the exception of eigenvalues zero
and $4 \Delta_\tau^2$, which have the multiplicity one. This was proved
previously for \nb{\kappa = 0} by H\"ogaasen-Feldman~\cite{Ho} and B\`es
and Broglia~\cite{BeBr}.

\Section {\label{cal}Calculations}

\Subsection {\label{eqd}Equidistant single-nucleon levels}

An insight into the general behavior of the model is obtained by
studying the idealized case of infinitely many equidistant
single-nucleon levels. This is the topic of the present section. More
precisely, I assume that the system has $d$ valence single-nucleon
levels $\epsilon_q$, spaced by a constant $D$ and symmetrically
distributed about zero, and that \nb{A_\va = 2 d}. The discussion in
Sect.~\ref{phs} then applies, and I denote by $\Delta$ the value of
$\Delta_\n$ for \nb{T = 0}. This system is considered in the limit $d
\to \infty$ with $D$, $\Delta$, and $\kappa$ kept fixed in the limit.
Appropriate values of $D$, $\Delta$, and $\kappa$ can be derived from
empirical formulas in the literature. Thus,
$D = 4 \big/ \bigl( 6 a / \pi^2 \bigr)$ with
$a = 0.176 A \bigl( 1 - 1.0 A ^{ -1/3 } \bigr) \MeV ^{ -1 }$ according
to Kataria, Ramamurthy, and Kapoor~\cite{KaRaKa}. An empirical formula
of Bohr and Mottelson~\cite{BoMo} for the odd-even mass difference gives
$\Delta = 12 A ^{ -1/2 } \MeV$. The constant $( D + \kappa ) / 2$ will
be seen to be close to the coefficient of $T(T+1)$ in a mass formula.
According to an empirical mass formula by Duflo and Zucker~\cite{DuZu},
then
$D + \kappa = 2 \bigl( 134.4 A ^{ -1 }- 203.6 A ^{ - 4/3 } \bigr) \MeV$.
Table~\ref{prm/ed}
\begin{table}
\caption{\label{prm/ed}
Parameters and results of the calculations with infinitely many
equidistant single-nucleon levels.}
\begin{ruledtabular}
\begin{tabular}{ccccccc}
$A$ & $D$ (MeV) & $\Delta / D$ & $\kappa / D$ & $a$ & $b$ & $x$ \\
\hline
\phz 48 &     1.1 & 1.6 & 2.0  & .873 & 4.08 & 1.010 \\
    100 & \phz .5 & 2.5 & 2.8  & .859 & 5.74 & 1.008 \\
    240 & \phz .2 & 4.2 & 3.6  & .842 & 8.71 & 1.005 \\
\end{tabular}
\end{ruledtabular}
\end{table}
shows, for some mass numbers $A$, the values of $D$, $\Delta / D$, and
$\kappa / D$ given by these expressions.

Since the dependence of $E$ on $\kappa$ is trivial according to
Eq.~\eqref{E}, I assume for now $\kappa = 0$. In units of $D$, the
energy $E$ is then a function of $\Delta / D$ and $T$. Some quantities
mentioned in the following are infinite in the limit $d \to \infty$.
When such quantities are said to obey some relation, this is meant to be
be arbitrarily accurately true when $d$ is sufficiently large. I have
checked that these statements hold very accurately in calculations for
$d = 500$ and $\Delta / D = 2.5$.

The Hartree-Bogolyubov energy $E_\HB$ is according to
Eq.~\eqref{EHB/fin} the sum of $E_0$ and $E_\pair$, both of which are
infinite in the limit $d \to \infty$. In this limit, the only change in
the solution of Eq.~\eqref{gap} and~\eqref{N/tau} when $N_\tau$ changes
by two units is a change of $\lambda_\tau$ by the level distance $D$.
Therefore, \nb{\Delta_\n = \Delta} independently of $T$, so $E_\pair$
does not contribute to the symmetry energy. On the other hand, $E_0$
does so. This is because the isospin $T$ is produced by the promotion of
$T$ nucleons from proton levels with the average energy $\lambda_\p / 2$
to neutron levels with the average energy $\lambda_\n / 2$. The result
is an increase
\begin{gather*}
  E_0 - E_{0,T = 0} = T \frac { \lambda_\n - \lambda_\p } 2
    = \hf D T ^2 \,,
\end{gather*}
where
\begin{gather}
  \frac { \lambda_\n - \lambda_\p } 2
    = \lambda_\n = - \lambda_\p = T \frac D 2 = \hf D T
  \label{lmn}
\end{gather}
has been used.

\begin{figure}
{\center \includegraphics{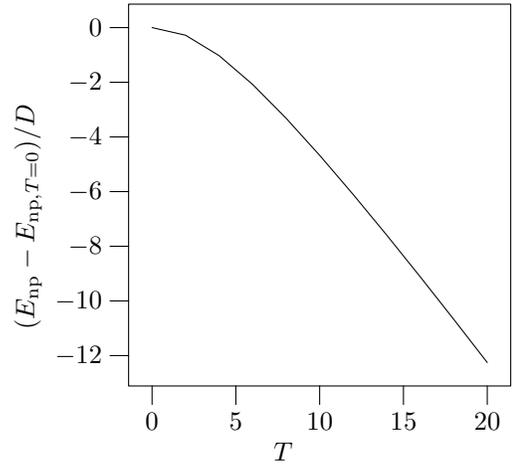} \par}
\caption{\label{fig:Enp}
The contribution of $E_\np$ to the symmetry energy in units of $D$ for
$\Delta / D = 2.5$ and $d = 1000$.}
\end{figure}

Eqs.~\eqref{ERPA/fin}, \eqref{Lms}, and~\eqref{lmn} give
\begin{gather}
  E_\RPA
    = E_\nn + E_\np + \hf D T
      + \sqrt { \bigl( \hf D T \bigr) ^2 + \Delta \rs ^2 } \,,
  \label{ERPA/ed} \\ \bk
  E_\nn
    = \sum_n \omega _{ \nn , n } - 2 \sum_q E_{ q \n } \,, \bk
  E_\np
    = \sum_n \omega _{ \np , n } - \sum_q E_{ q \n } \,,
  \nonumber
\end{gather}
where $\omega _{ \nn , n }$ are the positive eigenvalues of $\ma R$ in
the \emph{nn} or \emph{pp} space and $\omega _{ \np , n }$ the twofold
degenerate, positive eigenvalues of $\ma R$ in the \emph{np} space. By
adding the third term in the expression~\eqref{ERPA/ed} to $E_0$, one
gets a contribution to the symmetry energy proportional to $T(T+1)$,
\begin{gather*}
  E_0 - E_{0,T = 0} + \hf D T = \hf D T(T+1) \,.
\end{gather*}
The energies $E_\nn$ and $E_\np$ are infinite in the limit
$d \to \infty$. Because the quasinucleon energies $E_{ q \n }$ and the
elements of $\ma R$ in the \emph{nn} space are just relabeled when
$N_\n$ is changed by two units, $E_\nn$ is independent of $T$. On the
other hand, $E_\np$ gives a nonzero contribution to the symmetry energy.
This is demonstrated in Fig.~\ref{fig:Enp}
by a calculation for $\Delta / D = 2.5$ and $d = 1000$. A calculation
for $d = 500$ gives a curve which cannot be distinguished from this one.
The limit $d \to \infty$ is thus realized in practice for $d$ of this 
order.

The decrease of $E_\np$ with increasing $T$ is understood from the fact
noticed in connection with Eq.~\eqref{se} that the frequencies
$\omega _{ \np , n }$ lie between consecutive two-quasinucleon energies
\nb{E_{ q \n } + E_{ q \p } , \ q > 0}. The lowest two-quasinucleon
energy increases with $T$, whereas when $E_{ q \n } + E_{ q \p }$ is
sufficiently large, $E_{ q \n } + E_{ q \p } \approx
\nb { ( \epsilon_q - \lambda_\n ) + ( \epsilon_q - \lambda_\p ) }
= 2 \epsilon_q$, which is a constant. Therefore the difference between
the sum of $\omega _{ \np , n }$ and the sum of
\nb{E_{ q \n } + E_{ q \p } , \ q > 0}, that is, $\sum_q E_{ q \n }$,
decreases. The sum $E_\np + \sqrt { \Lambda_\pm }$ increases because all
the positive roots $\omega$ of Eq.~\eqref{se}, including
$\sqrt { \Lambda_\pm }$, and all the two-quasinucleon energies are
bounded below by $\sqrt { \Lambda_\pm }$, which increases. Therefore
also $E_\RPA$, which differs from the average of
$E_\np + \sqrt { \Lambda_\pm }$ only by the constant $E_\nn$, increases.

The calculation of $E_\np$ for $\Delta / D = 2.5$ and $d = 1000$ was
extended to larger $T$ than shown in Fig.~\ref{fig:Enp}, up to
\nb{T = 100}. The entire set of results is well decribed by the
expression
\begin{gather*}
  E_\np - E_{\np,T = 0}
    = - D \Bigl( \sqrt { ( a T ) ^2 + b ^2 } - b \Bigr) \,, \\
    a = .859 \,, \quad b = 5.74 \,.
  \label{Enp/emp}
\end{gather*}
In the view of Fig.~\ref{fig:Enp}, this gives a curve which can hardly
be distinguished from the calculated one. Similar results are obtained
for $\Delta / D = 1.6$ and $\Delta / D = 4.2$ with $a$ and $b$ given
in Table~\ref{prm/ed}.

Now collecting all the contributions discussed above and adding the 
last term in Eq.~\eqref{E}, one gets
\begin{gather}
  E - E_{ T = 0 } = \hf ( D + \kappa ) T(T+1) \bk
    - D \left( \sqrt { ( a T ) ^2 + b ^2 }
               - \sqrt { \biggl( \frac T 2 \biggr) ^2
               + \biggl( \frac { \Delta }  D \biggr) ^2 } 
    - b + \frac { \Delta } D \right) \,.
  \label{E/ed}
\end{gather}
For the paramters in Table~\ref{prm/ed}, the term subtracted from
\nb{\hf ( D + \kappa ) T(T+1)} in this expression is positive for 
$T > 0$. It amounts to 11\% of $\hf ( D + \kappa ) T$ for \nb{A = 48}
and \nb{T = 8} ($^{48}$S), 10\% for \nb{A = 100} and \nb{T = 14}
($^{100}$Kr), and 9\% for \nb{A = 240} and \nb{T= 28} ($^{240}$U). The
linear term $\hf ( D + \kappa ) T$ is thus the dominant correction to
the quadratic term $\hf ( D + \kappa ) T^2$. The sign of the additional
term in Eq.~\eqref{E/ed} is, however, consistent with the experience,
mentioned in Sect.~\ref{rev}, that in global fits to the empirical
masses with formulas which include a term proportional to $T(T+x)$, the
constant $x$ tends to be somewhat less than one. For \nb{T \approx 0},
Eq.~\eqref{E/ed} gives
\begin{gather}
  E - E_{ T = 0 } \approx \hf ( D + \kappa ) T(T+1)
    - D \biggl( \frac { a^2 } { 2 b }
    - \frac D { 8 \Delta } \biggr) T^2 \bk
  = \hf \Biggl(
      D \biggl( 1 - \frac { a^2 } b + \frac D { 4 \Delta }  \biggr)
      + \kappa \Biggr)
    T ( T + x ) \,, \bk
  x = \left( 1 - \frac { \ds \frac { a^2 } b - \frac D { 4 \Delta } }
                  { \ds 1 + \frac \kappa D } \right) ^{-1} \,.
  \label{x}
\end{gather}
Table~\ref{prm/ed} shows $x$ calculated from Eq.~\eqref{x} for the
parameters in the table. It is seen that $x \approx 1.01$ in these
cases.

As discussed in Sect.~\ref{npg}, the RPA gives the exact
energy~\eqref{E/noG} for $G = 0$. The symmetry energy derived from this
expression in the present case is
\begin{gather}
  E - E_{T = 0} = \hf \bigl( ( D + \kappa ) T ^2 + \kappa T \bigr) \,,
  \label{E/noG/ed}
\end{gather}
without the term $\hf D T$. The spontaneous breaking of the isobaric
invariance by the pairing force is thus required for this term to
appear.

\Subsection {\label{skm}Comparison with Skyrme force models}

Nuclear masses are often compared with Hartree-Fock (HF),
Hartree-Fock-BCS (HFBCS), or Hartree-Fock-Bogolyubov (HFB) calculations
with phenomenological energy functionals based on Skyrme forces. The HFB
method is described in detail in a recent article by Chamel, Goriely,
and Pearson~\cite{ChGoPe}, where also references to earlier work in this
line of research are found. Applying the HF, HFBCS or HFB scheme in an
approximate way to the present Hamiltonian sheds a light on the origin
of certain phenomena observed in the Skyrme force calculations. In the
formalism of Ref.~\cite{ChGoPe}, a pairing and a particle-hole part of
the two-nucleon interaction are treated differently, as in the present
theory. For the present pairing interaction and $\av { P_z } = 0$,
there is then no difference between the HFBCS and HFB schemes.

\begin{figure}
{\center \includegraphics{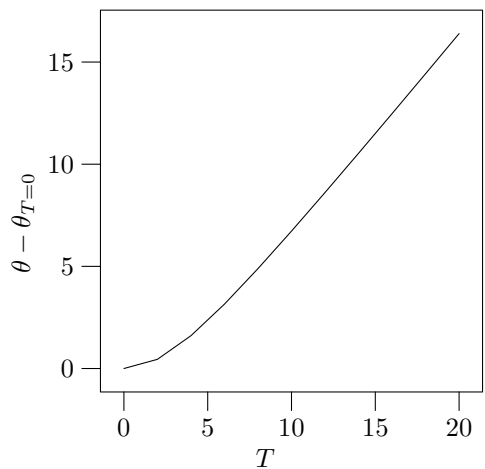} \par}
\caption{ \label{fig:theta}
The quantity $\theta - \theta_{ T = 0 }$ as a function of $T$ for
$d = 1000$ and $\Delta / D = 2.5$. }
\end{figure}

Apart from the different interactions, the only difference between the
formalism of Sect.~\ref{vac} and the HFB formalism of Ref.~\cite{ChGoPe}
is that the particle-hole matrix element is antisymmetrized in the
latter. In the formalism of Sect.~\ref{vac}, this amounts to replacing
$R_\HB$ with
\begin{gather}
  R_\HB - \hf \kappa \,\tr \, \vc t \cdot \rho \vc t \rho \,, \quad
    \me j \rho k = \av { \hc a k \ba a j } \,.
  \label{RHFB}
\end{gather}
The resulting Routhian is stationary at the quasinucleon vacuum
$\ke \Phi$ given by Eqs.~\eqref{Phi}, \eqref{alpha}, \eqref{uv},
\eqref{Eq}, \eqref{gap}, and \eqref{N/tau} when $\epsilon_q$ is replaced
with $\epsilon_q
- \frac 1 4 \kappa ( v _{ q \tau } ^2 + 2 v _{ q \tau' } ^2 )$,
$\tau' \ne \tau$. I neglect this modification of the self-consistent
single-nucleon energy and calculate the second term in Eq.~\eqref{RHFB},
the ``Fock term'', with $u_{ q \tau }$ and $v_{ q \tau}$ given by
Eqs.~\eqref{uv} and~\eqref{Eq}. The essential factor is
\begin{gather}
  \tr \, \vc t \cdot \rho \vc t \rho
    = \sum_q \Bigl( \hf \bigl( v_{ q \nÊ} ^4 + v_{ q \pÊ} ^4 \bigr)
                    + 2 v_{ q \nÊ} ^2 v_{ q \pÊ} ^2 \Bigr) \bk
    = \tfrac 3 4 A_\va - \sum_q \Bigl(
        \hf \bigl( u_{ q \nÊ} ^2 v_{ q \nÊ} ^2
                   + u_{ q \pÊ} ^2 v_{ q \pÊ} ^2 \bigr)
        + u_{ q \nÊ} ^2 v_{ q \pÊ} ^2
        + u_{ q \pÊ} ^2 v_{ q \nÊ} ^2 \Bigr) \,.
  \label{exch}
\end{gather}
The first term in this expression gives the second term in the outer
bracket in Eq.~\eqref{c} and the second term a part of the first of the
diagrams~\eqref{Rgr}. In the case of infinitely many equidistant
single-nucleon levels, the only part of the expression~\eqref{exch}
which depends on $T$ is
\begin{gather}
  \theta = \sum_q
    \bigl( u_{ q \nÊ} ^2 v_{ q \pÊ} ^2
         + u_{ q \pÊ} ^2 v_{ q \nÊ} ^2 \bigr) \,. 
  \label{theta}
\end{gather}
The quantity $\theta - \theta_{ T = 0 }$ is shown in
Fig.~\ref{fig:theta}
as a function of $T$ for $d = 1000$ and $\Delta / D = 2.5$. It is seen
that the asymtotic slope is one. This is easily understood from
Eq.~\eqref{theta}. Indeed, for $T \gg 1$ the first term in the bracket
is practically zero, while the second term is close to one for
$\lambda_\p < \epsilon_q < \lambda_\n$ and close to zero otherwise.
Hence, for $T \gg 1$, the contribution of the Fock term to the symmetry
energy deviates from the linear term in the expression~\eqref{E/noG/ed}
only by a constant. For $T \approx 0$, it is, however, quadratic in $T$
because $\theta$ is by Eqs.~\eqref{uv}, \eqref{Eq}, \eqref{lmn},
and~\eqref{theta} analytic and even as a function of $T$. Therefore, the
HFB scheme does not produce any cusp at $N = Z$ in the curve of masses
along an isobaric chain. The study by Satu\l a and Wyss~\cite{SaWySym}
shows that this is true also in HFBCS calculations with Skyrme forces,
and it is true in the HFB calculations discussed in Ref.~\cite{ChGoPe},
where, in order to reproduce the empirical masses, a so-called Wigner
correction is added to the HFB energy.

HF calculations with Skyrme forces correpond to the case $G = 0$ of the
present theory. I this case, the Fock term is equal to
$\hf \kappa ( T - \tfrac 3 4 A_\va)$, so the exact energy~\eqref{E/noG}
is recovered. With infinitely many equidistant single-nucleon levels,
the symmetry energy is given by Eq.~\eqref{E/noG/ed} and thus includes
the linear term $\hf \kappa T$. Satu\l a and Wyss~\cite{SaWySym} find
that such a term appears also in HF calculations with Skyrme forces. As
in the present theory, there is in these calculations no term
corresponding to $\hf D T$. The present schematic model thus seems
representative of the Skyrme force models as to the basic mechanisms
responsible for the absense or presense of such linear terms of various
origins.

\Subsection {\label{def}Deformed Woods-Saxon single-nucleon levels}

Moving now from the idealized case of a uniform single-nucleon spectrum
to assumptions closer to the reality, I discuss in this section
calculations with single-nucleon levels derived from a deformed
Woods-Saxon potential. I have chosen for this study isobaric chains
whose $T = 0$ members have low-lying $2^+$ levels and can thus be
supposed to have appreciable quadrupole deformations. For each isobaric
chain, the single-nucleon energies are calculated for a deformation
pertaining to the $T = 0$ nucleus. The definition of the deformed
Woods-Saxon potential follows Dudek~\etal~\cite{DuMa*}. The
average nucleon mass and the average of the ``universal'' neutron and
proton parameters of Dudek~\etal~\cite{DuSz*} are employed, and no
Coulomb potential is included. The halfdepth surface of the
spin-independent part of the potential is assumed to have a pure,
prolate quadrupole shape with the deformation $\beta$ derived by the
relations of Raman, Nestor, and Tikkanen~\cite{RaNeTi} from
$B(E2,0^+ \to 2^+)$, when it is known, and else from the $2^+$
excitation energy. All bound single-nucleon states are included in the
valence space.

The pairing force contant $G$ is determined for each isobaric chain by
demanding $\Delta_\n = \Delta_\p = 12 A^{-1/2}\MeV$ for $T = 0$. Most
other nuclei in the isobaric chain then acquire similar values of
$\Delta_\tau$, but $\Delta_\tau$ may vanish if $\lambda_\tau$ reaches a
gap in the single-nucleon spectrum. The calculated symmetry energy turns
out to be minor sensitive to the precise value of~$G$. The constant
$\kappa$ is chosen so as to best fit the data. A~summary of the 
resulting parameters is given in the first, third, and fourth row of
Table~\ref{prm/ws}.
\begin{table}
\caption{\label{prm/ws}
Parameters and results of the calculation with Woods-Saxon
single-nucleon levels.}
\begin{ruledtabular}
\begin{tabular}{ccccc}
$A$ & $\beta$ & $G$ (MeV) & Interval of $\Delta_\tau$ (MeV) &
$\kappa$ (MeV) \\ \hline
\phz 48 & \phz .337 & .450     & 1.62--1.85 &     1.4 \\
\phz 56 & 0\phd     & .4\phzz  &   0--1.94  &     1.1 \\
\phz 68 & \phz .234 & .286     &   0--1.65  &     1.5 \\
\phz 80 & \phz .342 & .262     & 1.08--1.69 &     1.1 \\
    100 & 0\phd     & .2\phzz  &   0--1.55  & \phz .8
\end{tabular}
\end{ruledtabular}
\end{table}
A somewhat disturbing feature of these parameters is the slightly
irregular $A$ dependence of the optimal~$\kappa$. This irregularity may
be related to the crude treatment of the shape degrees of freedom.

For comparison with the calculated results, I have derived an empirical
symmetry energy from the masses compiled by Audi, Wapstra, and Thibault
in their 2003 Atomic Mass Evaluation~\cite{AuWaTh}. It is calculated
from the binding energy for $M_T \ge 0$ minus the electrostatic term
extracted by Kirson~\cite{KiGro} from the differences of the binding
energies of mirror nuclei,
\begin{gather*}
  \frac { - .716 Z^2 + .993 Z^{4/3} } { A^{1/3} } \MeV \,.
\end{gather*}

\begin{figure}
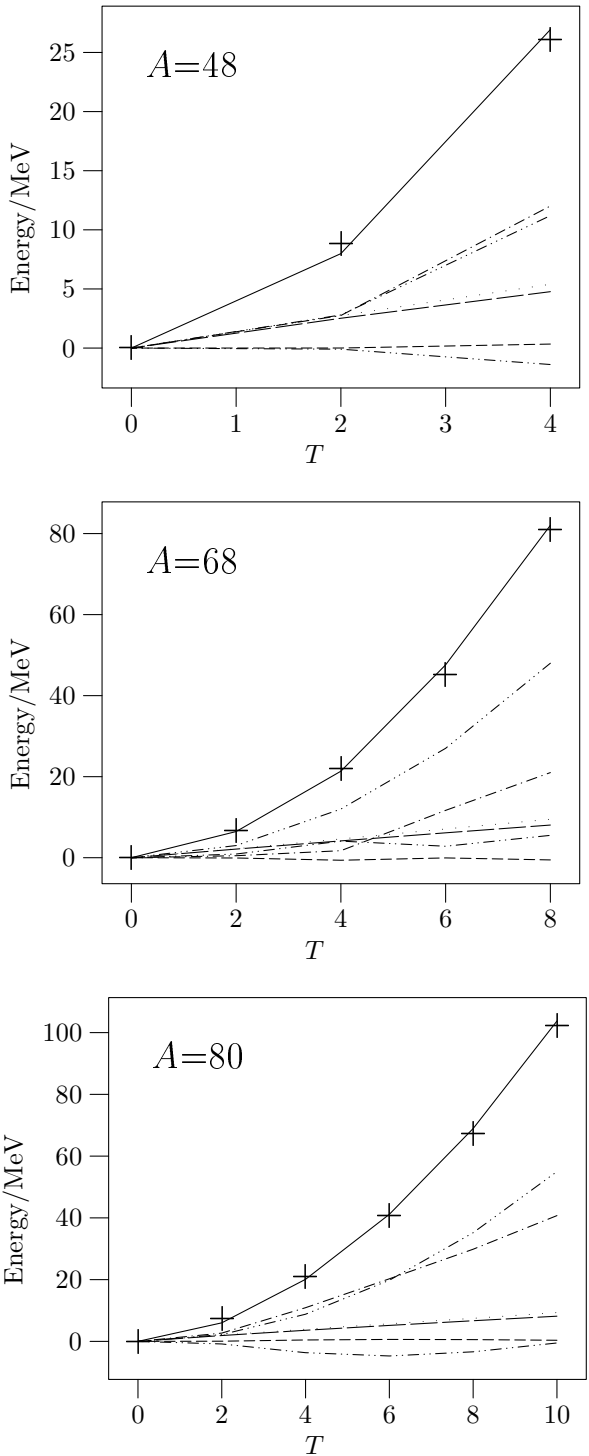

{\center \includegraphics{48.mps} \par}
{\center \includegraphics{68.mps} \par}
{\center \includegraphics{80.mps} \par}
\caption{ \label{fig:def} Calculated (solid line) and empirical
(crosses) symmetry energy for some values of the mass number $A$. Shown
with different signatures are also the following components of the
calculated symmetry energy: \nb{E_0 - E_{ 0 , T = 0 }} (dash-dot),
\nb{E_\pair - E_{ \pair , T = 0 } } (dash-dot-dot), $\hf \kappa T^2$
(dash-dot-dot-dot),
\nb{E_{ \RPA , \nnpp } - E_{ \RPA , \nnpp, T = 0 }} (short dash),
\nb{E_{ \RPA , \np } - E_{ \RPA , \np, T = 0 }} (long dash), where
$E_{ \RPA , \nnpp }$ and $E_{ \RPA , \np }$ are the contributions to the
expression~\eqref{ERPA/fin} from the \emph{nn}+\emph{pp} and \emph{np}
spaces. The dotted curve is $\hf \mu$.}
\end{figure}

The calculated and empirical symmetry
energies are shown in Fig.~\ref{fig:def}.
In view of the schematic character of the model, where, in particular,
any variation of shape degrees of freedom is absent, the agreement is
suprisingly good. Also shown in the figure is the composition of the
calculated symmetry energy. The term $\hf \kappa T ^2$ typically makes
up about half of it for large $T$. The contribution of the pairing
energy $E_\pair$ is plus or minus a few MeV. It arises from the
variation of $\Delta_\tau$ along the isobaric chain. This variation is
due, in turn, to the variation with $\lambda_\tau$ of the local
single-nucleon level density. The irregularity of the single-nucleon
spectrum is also responsible for a fairly irregular behavior of the
contribution from $E_0$ as compared to its quadratic dependence on $T$
in the idealized case.

The part of the contribution from $E_\RPA$ originating in the
\emph{nn}+\emph{pp} space is essentially zero, as in the idealized case.
The part originating in the \emph{np} space is dominated by the single
eigenvalue $\hf \mu$ of $\ma R$, which increases essentially linearly
with $T$. Its deviation from $\hf \mu$ appears to be of higher than
linear order in $T$ and always negative, just as in the idealized case,
and it amounts to about $- 1.5 \MeV$ for the largest $T$'s for which the
binding energy is measured. The ratio $\nb{ ( E_\RPA - E_{ \RPA , T = 0
} ) } / \nb{ ( E - E_{ T = 0 } ) }$ is close to \nb{ T / ( T^2 + T ) =
1/3} for $T = 2$ in the three cases. This is understood from the
following facts. (1)~$\nb{E_\HB - E_{ \HB , \av { T_z } = 0 } } \propto
\av { T_z } ^2$ for small $\av { T_z }$ because $E_\HB$ is analytic and
even as a funtion of $\av { T_z }$. (2)~$E_\RPA - E_{ \RPA , T = 0 }$ is
dominated by \nb{\hf \mu = \hf d E_\HB / d \av { T_z }}. (3)~$\av { T_z
} = T$.

\Subsection {\label{sph}Spherical Woods-Saxon single-nucleon levels:
             yet another mechanism}

\begin{figure}
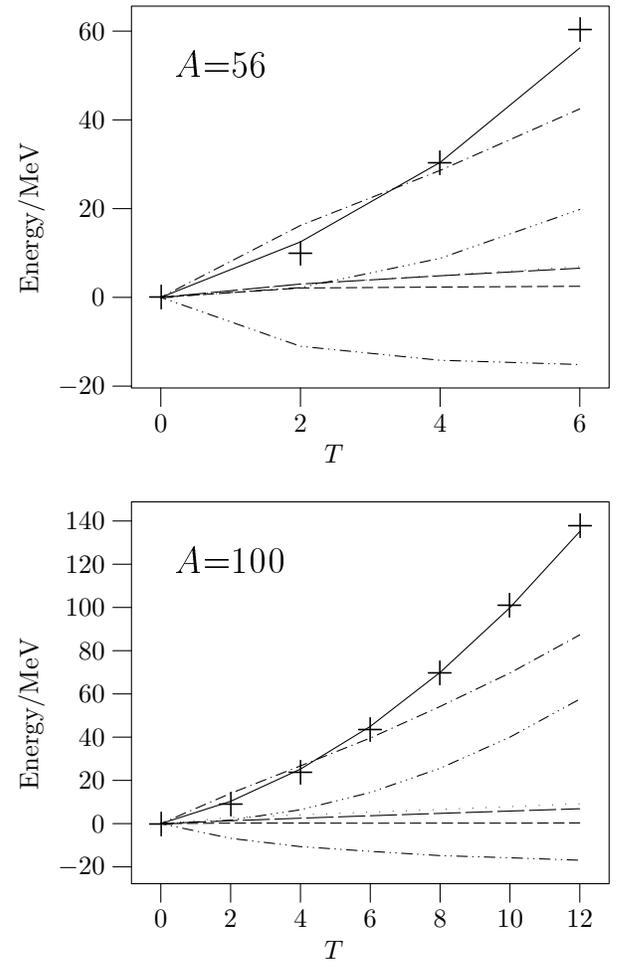

{\center \includegraphics{56.mps} \par}
{\center \includegraphics{100.mps} \par}
\caption{ \label{fig:sph} See the caption to Fig.~\ref{fig:def}.}
\end{figure}

For $A = 56$ and $A = 100$, the $T = 0$ nucleus is doubly magic.
Therefore, I have made calculations for these mass numbers with a
spherical Woods-Saxon potential. I have chosen $G$ in this case so
that $\Delta_\tau \approx 12 A^{-1/2}\MeV$ for $T > 0$. Then,
$\Delta_\tau = 0$ for $T = 0$. The adopted parameters are given in the
second and fifth row of Table~\ref{prm/ws}, and the results of the
calculations are shown in Fig.~\ref{fig:sph}.
The valence space consists of the levels below the $N_\tau = 50$ shell
closure for $A = 56$, and below the $N_\tau = 82$ shell closure for
$A = 100$.

The contributions to the symmetry energy from $\hf \kappa T^2$ and
$E_\RPA$ are similar to those of the deformed case. But, due to the
increase of $\Delta_\tau$ when $\lambda_\tau$ moves into a shell above
or below the magic gap, $E_\pair$ gives in this case a significant
negative contribution. Because the largest decrease of $E_\pair$ takes
place for low $T$, $E_\pair$ then contributes to the nonlinear part of
the symmetry energy. On the other hand, $E_0$ is practically linear.
This is because the increase of $E_0$ results from the promotion of
nucleons from proton states below the shell gap to neutron states above
the gap. The slope of $E_0$ is thus roughly equal to the shell gap. The
slope of $E_0 + E_\pair$ at low $T$ is similar to that of $E_\RPA$ for
\nb{A = 56} and several times that of $E_\RPA$ for \nb{A = 100}. In
both cases, it thus gives a significant contribution to the linear part
of the symmetry energy. This could be expected to be true in general in
isobaric chains with a doubly magic $T = 0$ nucleus.

\Section {\label{sig}Experimental signature of superfluid isorotation}

An experimental signature of a nuclear rotation in space is the sequence
of large and approximately equal reduced $E2$ transition probabilities
between consecutive members of a rotational band produced by the
quadrupole deformation of the intrinsic charge distribution. The
quadrupole deformation is measured by the mass or charge quadrupole
tensor. In the case of superfluid isorotation, the intrinsic deformation
is measured by the isovector $\vc P$. Since the consecutive members of
an isorotational band are separated by two units of isospin, they are
not connected directly by an isovector. However, as pointed out by Bohr
and Mottelson~\cite{BoMo}, the superfluid isorotational bands
participate in a larger structure consisting of the ground states of all
the doubly even nuclei and their isobaric analog states. The ground
states of the doubly even isotopes are connected by $P_\n$ and the
ground states of the doubly even isotones by $P_\p$. The isovector
component $P_z$ connects such states with \nb{\ab{ M_T } = T - 1}
isobaric analog states in doubly odd nuclei. The chains of superfluid
isotopes or isotones in fact form the pair rotational bands discussed by
Bohr~\cite{Bo}. Yoshida~\cite{Yo} shows that superfluidity enhances the
ground state to ground state cross section of two-neutron or two-proton
transfer between doubly even nuclei by a factor about
$(2 \Delta_\tau / G ) ^2$, where \nb{\tau = \n} or~$\p$. Isovector
one-neutron-one-proton transfer between a ground state of a doubly even
nucleus and an isobaric analog state is then similarly enhanced.

The experimental signature of superfluid isorotation therefore coincides
with that of pair rotation. The picture of a superfluid isorotation
implies that the enhancement factors of two-nucleon transfer involving
doubly even ground states or their isobaric analogs remains
approximately constant all the way down to \nb{T = 0}. Near closed
shells, the pair rotational bands may develop into the pair vibrational
bands dicussed by Bohr~\cite{Bo}, which, as well, have enhanced
two-nucleon transfer cross sections. If, on the other hand, doubly even
nuclei with \nb{T = 0} would have a structure radically different from
that of doubly even nuclei with \nb{T > 0}, a major deviation from such
a smooth behavior would be seen. No such effect seems to be indicated by
the enhancement factors compiled by B\`es~\etal~\cite{Be*}.

\Section {\label{sum}Summary}

A Hamiltonian with a single-nucleon term and a two-nucleon interaction
was investigated. The single-nucleon Hamiltonian has fourfold degenerate
eigenvalues correponding to time reversed pairs of neutron and proton
states. The two-nucleon interaction has a pairing and a particle-hole
part. The paring part is the isobarically invariant isovector pairing
force and the particle-hole part an interaction of isospins, which I
call the symmetry force. A Routhian was contructed by subtracting from
the Hartree-Bogolyubov energy functional Lagrangian multiplier terms
propotional to the expectation values of the number of valence nucleons
and the $z$ component of the isospin, and it was shown that this
Routhian is locally minimized by a product of neutron and proton
Bardeen-Cooper-Schrieffer~\cite{BaCoSc} states.

This quasinucleon vacuum and a single-quasinucleon Routhian operator
derived from the Hartree-Bogolyubov Routhian was taken as the starting
point for a calculation of the ground state energy as a function of the
number of valence particles and the isospin quantum number $T$ in the
Random Phase Approximation (RPA). The correction to the
Hartree-Bogolyubov energy is the sum of a term which does not depend on
$T$ and a term $E_\RPA$ equal to half the sum of the poles of the RPA
two-quasinucleon propagator minus the sum of the two-quasinucleon
energies. The poles of the two-quasinucleon propagator which are
different from two-quasinucleon energies can be determined separately in
a two-neutron, a two-proton, and a neutron-proton quasiparticle space.
In each of these spaces, there is a Nambu-Goldstone pole due to the
global gauge invariance and isobaric invariance of the Hamiltonian. The
two-neutron and two-proton Nambu-Goldstone poles have the frequency
zero, while the neutron-proton Nambu-Goldstone pole is equal to the
Lagrangian multiplier of the $z$ component of the isospin. The term in
$E_\RPA$ resulting from this pole was interpreted in a picture of a
collective rotation in isospace to be due to the quantal fluctuation of
the isospin. The pole in question is the only one which depends on the
strength $\kappa$ of the symmetry force. The only contribution of the
symmetry force to the symmetry energy is therefore a term
$\hf \kappa T(T+1)$.

If the single-nucleon spectrum is symmetric about a certain energy and
the valence space halfway filled, the neutron-proton poles are twofold
degenerate except for the Nambu-Goldstone pole and one more pole, which
has an analytic expression. Related results are known from the
literature to pertain to the neutron-neutron and proton-proton spaces.

In an idealized case of infinitely many equidistant single-nucleon
levels, the neutron and proton systems have a common pair gap $\Delta$
which does not depend on $T$. Therefore, the pairing force does not
contribute to the symmetry energy. Neither do the two-neutron and
two-proton parts of $E_\RPA$. For \nb{\kappa = 0}, the single-nucleon
term in the Hamiltonian and the neutron-proton Nambu-Goldstone pole give
together a contribution equal to $\hf D T(T+1)$, where $D$ is the
single-nucleon level spacing. The second nondegenerate neutron-proton
pole gives a contribution equal to
$\sqrt { \bigl( \hf D T \bigr) ^2 + \Delta \rs ^2 } - \Delta$. The
remainder of the contribution of the neutron-proton part of $E_\RPA$ was
calculated numerically. In a very good approximation, it is
$- D \Bigl( \sqrt { ( a T ) ^2 + b ^2 } - b \Bigr)$, where $a$ and $b$
are functions of $\Delta / D$. For realistic parameters, the sum of
these two terms is negative and amounts to about $- 10 \%$ of the linear
term $\hf ( D + \kappa ) T$ for the largest $T$ of observed nuclei.
For \nb{T \approx 0}, they give a contribution to the symmetry energy
quadratic in $T$ which makes the symmetry energy proportional to
$T(T+x)$ with $x \approx 1.01$. In the absence of the pairing force, the
RPA gives the exact symmetry energy, which is
$\hf \bigl( ( D + \kappa ) T ^2 + \kappa T \bigr)$. This expression does
not have the linear term $\hf D T$, which thus appears only when the
isobaric invariance is spontaneously broken by the pairing force.

If the matrix element of the symmetry force is antisymmetrized and the
contribution of the exchange term to the self-consistent single-nucleon
energy and the RPA correlations are neglected, the total contribution of
the symmetry force to the symmetry energy is asymptotically for large
$T$ equal to $\hf \kappa T(T+1)$ plus a constant. For \nb{T \approx 0},
it is quadratic in $T$. Therefore the curves of masses along isobaric
chains get no cusps at \nb{T = 0}. This corresponds to observations
reported from Hartree-Fock-Bogolyubov calculations with Skyrme forces.
In the absence of the pairing force, the exact symmetry energy without
the linear term $\hf D T$ is recovered also in this approximation. This
correponds to observations reported from Skyrme force Hartree-Fock
calculations.

Calculations with Woods-Saxon single-nucleon levels give results in
surprisingly good agreement with the empirical variation of the binding
energy of doubly even nuclei along isobaric chains. In the case of a
deformed Woods-Saxon potential, the behavior of the individual
components of the calculated symmetry energy is similar to their behavor
in the idealized case. In calculations for \nb{A = 56} and \nb{A = 100}
with spherical Woods-Saxon levels, the promotion of nucleons across the
\nb{N = Z = 28} and \nb{N = Z = 50} gaps in the single-nucleon spectrum
gives a large linear contribution. Due to the onset of superfluidity
when the neutron and proton Fermi levels move into the shells around
these gaps, the pairing force gives another large contibution in this
case. Together, these two contributions give a linear term which is
comparable to or larger than that of $E_\RPA$.

In a picture of a collective rotation in isospace, the intrinsic
deformation is measured by the pair annihilation isovector. The
isorotation can therefore be characterized as superfluid. The pair
annihilation isovector does not connect directly consecutive members of
a superfluid isorotational band, which differ by two units of isospin.
However, the superfluid isorotational bands participate in a larger
structure which includes the pair rotational and pair vibrational bands
discussed in the literature. Within these bands, the cross sections for
two-nucleon transfer are enhanced by the superfluid correlations. The
picture of a superfluid isorotation implies that the enhancement factors
of isovector two-nucleon transfer should develop smoothly down to
\nb{T = 0}. This seems to be consistent with the empirical evidence.

\begin{acknowledgments}

I thank Friedrich D\"onau for providing me with the Woods-Saxon code.

\end{acknowledgments}

\appendix*

\Section {\label{SU2}Wigner argument for the isobaric $SU(2)$}

Consider a system of valence nucleons and a two-nucleon interaction
$\sum v$, where the summation is over pairs of nucleons, and let $P_\sy$
and $P_\ay $ project to the spaces of two-nucleon states symmetric and
antisymmetric in position and spin. In any state of the system,
\begin{gather}
  \Av {\ts \sum v} = \Av {\ts \sum ( P_\sy + P_\ay ) v} \bk
    = \frac {\Av {\sum P_\sy v}} {\Av {\sum P_\sy}}
        \Av {\ts \sum P_\sy}
    + \frac {\Av {\sum P_\ay v}} {\Av {\sum P_\ay}}
        \Av {\ts \sum P_\ay} \,.
  \label{Wig}
\end{gather}
The total antisymmetry implies
\nb{P_\sy = \frac14 - \vc t_1 \cdot \vc t_2}, where $\vc t_1$ and
$\vc t_2$ are the isospins of the two nucleons. Hence
\begin{gather*}
  {\ts \sum P_\sy} = \frac {\hat A_\va ( \hat A_\va + 2 ) } 8
    - \frac { \vc T ^2 } 2 \,, \\
  {\ts \sum P_\ay} = {\ts \sum ( 1 - P_\sy ) } 
     = \frac {3 \hat A_\va ( \hat A_\va - 2 ) } 8
       + \frac { \vc T ^2 } 2 \,.
\end{gather*}
The assumption in the case of the isobaric $SU(2)$ symmetry
corresponding to Wigner's in the case of the $SU(4)$ symmetry is that
the ground state values of $\Av {\sum P_\sy v}/\Av {\sum P_\sy}$ and
$\Av {\sum P_\ay v}/\Av {\sum P_\ay }$ are smooth functions of $A_\va$
and $T$. If, in particular, they do not depend on $T$, the symmetry
energy is proportional to $T(T+1)$.

The assumption of $T$-independent
$\Av {\sum P_\sy v}/\Av {\sum P_\sy}$ and
$\Av {\sum P_\ay v}/\Av {\sum P_\ay }$ is easily seen to be invalid in
the case of the isobarically invariant isovector pairing force acting in
a single $j$-shell. In this case, \nb{v = - (2j+1) G P_0}, where $P_0$
projects to two-nucleon angular momentum zero. From an expression for
the eigenvalues of \nb{(2j+1) \sum P_0} derived by Edmonds and
Flowers~\cite{EdFl}, it follows that the lowest eigenvalue of $\sum v$
for fixed even $A_\va$ and $A_\va / 2 + T$ is
\begin{gather}
  G \biggl( T(T+1) - \frac {A_\va ( 4j+8 - A_\va ) } 4 \biggr) \,.
  \label{EF}
\end{gather}
Since a state of two nucleons from the same $j$-shell with angular
momentum zero is antisymmetric in position and spin, and $v$ is negative
semidefinite, \nb{\Av {\sum P_\sy v} = 0} and
\nb{\Av {\sum P_\ay v} < 0} unless \nb{\Av {\sum v} = 0}. Taking
$\Av {\sum P_\ay v} / \Av {\sum P_\ay }$ to be a negative constant, one
gets from Eq.~\eqref{Wig}
\begin{gather*}
  \Av {\ts \sum v} \propto \Av {\ts \sum P_\ay}
    = \frac {3A_\va(A_\va-2)} 8 + \frac {T(T+1)} 2
\end{gather*}
with a negative constant of proportionality. This expression obviously
conflicts with the expression~\eqref{EF}. It even differs from the
latter by the sign of its contribution to the symmetry energy.

\bibliography{pairsym}

\end{document}